%% file: main.tex
\pdfoutput=1
\documentclass[12pt,a4paper]{article}

\usepackage{ifthen} 
\newboolean{pdflatex}
\setboolean{pdflatex}{true} 

\newboolean{articletitles}
\setboolean{articletitles}{true} 

\newboolean{uprightparticles}
\setboolean{uprightparticles}{false} 

\usepackage{booktabs} 
\usepackage{array}

\def\paperauthors{} 
\def\paperasciititle{Performance of the LHCb muon detector in Run 3} 
\def\papertitle{Performance of the \lhcb muon detector in Run 3} 
\def\papercopyright{\the\year\ CERN for the benefit of the LHCb collaboration} 
\def\paperlicence{CC BY 4.0 licence}
\def\paperlicenceurl{https://creativecommons.org/licenses/by/4.0/}

\newif\ifEnableSectionTOCLinks
\EnableSectionTOCLinksfalse 

\usepackage[top=1in, bottom=1.25in, left=1in, right=1in]{geometry}

\usepackage{microtype}
\usepackage{lineno}  
\usepackage{xspace} 
\usepackage{caption} 

\usepackage{graphicx}  
\usepackage{color}
\usepackage{colortbl}
\graphicspath{{./figs/}} 

\usepackage{amsmath} 
\usepackage{amssymb}
\usepackage{amsfonts}
\usepackage{upgreek} 

\newcommand*\patchAmsMathEnvironmentForLineno[1]{%
\expandafter\let\csname old#1\expandafter\endcsname\csname #1\endcsname
\expandafter\let\csname oldend#1\expandafter\endcsname\csname
end#1\endcsname
 \renewenvironment{#1}%
   {\linenomath\csname old#1\endcsname}%
   {\csname oldend#1\endcsname\endlinenomath}%
}
\newcommand*\patchBothAmsMathEnvironmentsForLineno[1]{%
  \patchAmsMathEnvironmentForLineno{#1}%
  \patchAmsMathEnvironmentForLineno{#1*}%
}
\AtBeginDocument{%
\patchBothAmsMathEnvironmentsForLineno{equation}%
\patchBothAmsMathEnvironmentsForLineno{align}%
\patchBothAmsMathEnvironmentsForLineno{flalign}%
\patchBothAmsMathEnvironmentsForLineno{alignat}%
\patchBothAmsMathEnvironmentsForLineno{gather}%
\patchBothAmsMathEnvironmentsForLineno{multline}%
\patchBothAmsMathEnvironmentsForLineno{eqnarray}%
}

\usepackage[pdftex,
    pdftitle={\paperasciititle},
    pdfauthor={\paperauthors}
]{hyperref}

\usepackage[colorinlistoftodos,textsize=scriptsize]{todonotes}

\usepackage[bottom,flushmargin,hang,multiple]{footmisc}

\usepackage[all]{hypcap} 

\input{symbols}

\hypersetup{
  colorlinks   = true, 
  urlcolor     = blue, 
  linkcolor    = blue, 
  citecolor    = red   
}

\ifEnableSectionTOCLinks
    \usepackage[explicit]{titlesec} 
    
    \let\oldcontentsline\contentsline
    \renewcommand

    \titleformat{\section}{\normalfont\Large\bf}{\hyperlink{tocsection.\thesection}{{\thesection} \parbox[t]{\dimexpr\textwidth-1pc}{#1}}}{1pc}{}

    \titleformat{\subsection}{\normalfont\bf}{\hyperlink{tocsubsection.\thesubsection}{{\thesubsection} \parbox[t]{\dimexpr\textwidth-1pc}{#1}}}{1pc}{}

    \titleformat{name=\section,numberless}[display]{}{}{0pt}{\normalfont\Huge\bfseries #1}
\fi

\usepackage{cite} 
\usepackage{mciteplus}

\usepackage{longtable} 
\usepackage{makecell}

\begin{document}

\renewcommand{\thefootnote}{\fnsymbol{footnote}}
\setcounter{footnote}{1}

\begin{titlepage}
\pagenumbering{roman}

\vspace*{-1.5cm}
\centerline{\large EUROPEAN ORGANIZATION FOR NUCLEAR RESEARCH (CERN)}
\vspace*{1.5cm}
\noindent
\begin{tabular*}{\linewidth}{lc@{\extracolsep{\fill}}r@{\extracolsep{0pt}}}
\vspace*{-1.5cm}\mbox{\!\!\!\includegraphics[width=.14\textwidth]{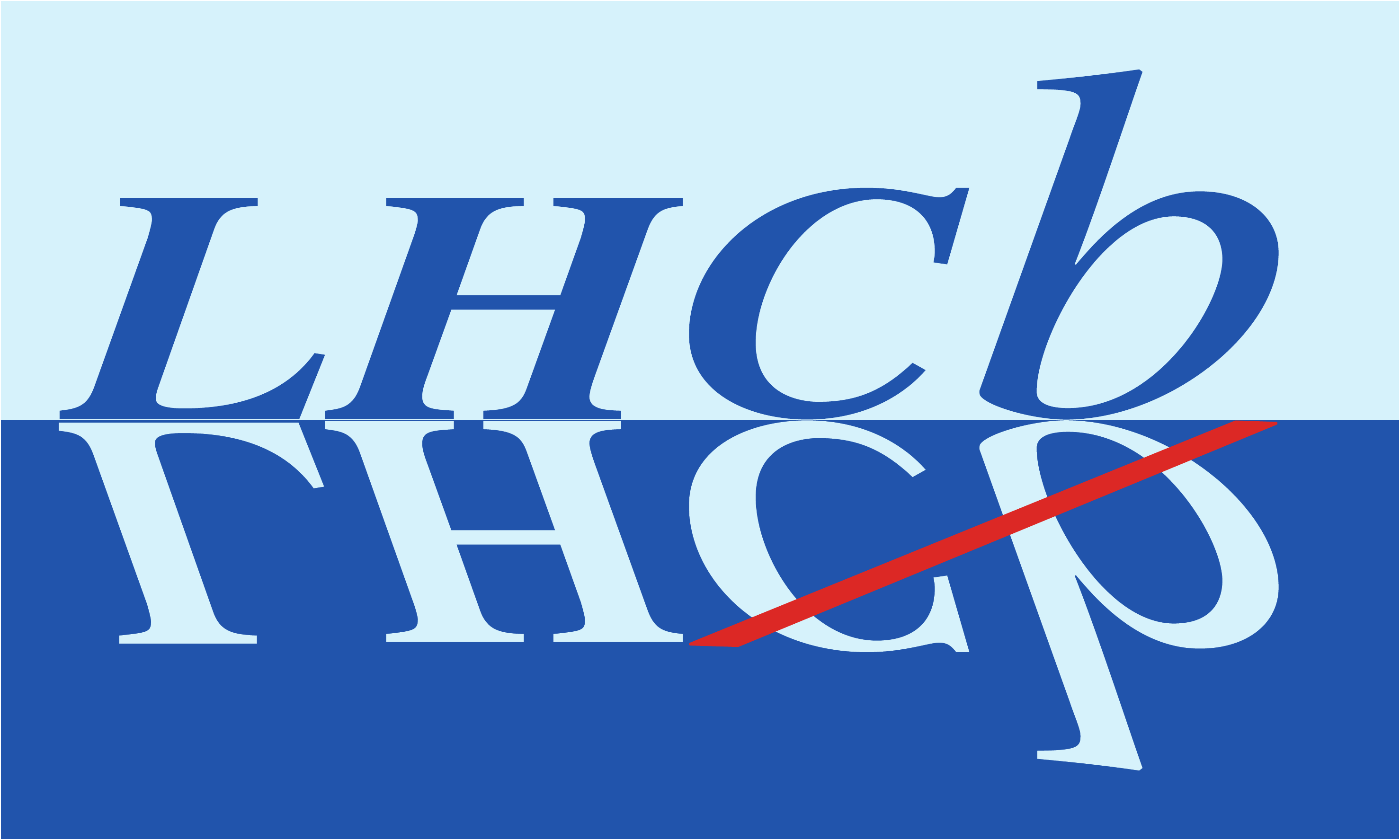}} & &
\\
 & & CERN-LHCb-DP-2025-007 \\  
 & & \today \\ 
 & & \\
\end{tabular*}

\vspace*{0.4cm}

{\normalfont\bfseries\boldmath\huge
\begin{center}
  \papertitle 
\end{center}
}

\vspace*{0.4cm}

\input{authors.tex}

\begin{abstract}
  \noindent
In Run 3 of the LHC, the instantaneous luminosity at the LHCb interaction point has been increased by a factor of five, from $\mathcal{L}=4\times 10^{32}\cm^{-2}\sec^{-1}$ to $\mathcal{L}=2\times 10^{33}\cm^{-2}\sec^{-1}$. 
Several hardware interventions, including a complete overhaul of the readout electronics, have been carried out on the muon detector. The muon identification algorithms in the software trigger were improved with the aim of ensuring Run 2 performance under a higher particle rate. The operation and calibration of the upgraded muon detector are presented. The muon detection efficiency and muon identification performance are evaluated on data calibration samples collected during the year 2024. A muon identification efficiency above 90\% with sub-percent hadron misidentification probability is achieved by exploiting the pattern of hits in the muon detector.

\end{abstract}

\vspace*{2.0cm}

\begin{center}
Published in Nucl. Instrum. Meth. A 1089 (2026) 171588
\end{center}

\vspace{\fill}

{\footnotesize 
\centerline{\copyright~\papercopyright. \href{\paperlicenceurl}{\paperlicence}.}}
\vspace*{2mm}

\end{titlepage}


\newpage
\setcounter{page}{2}
\mbox{~}


\renewcommand{\thefootnote}{\arabic{footnote}}
\setcounter{footnote}{0}

\tableofcontents

\cleardoublepage


\pagestyle{plain} 
\setcounter{page}{1}
\pagenumbering{arabic}



\section{Introduction}
\label{sec:Introduction}

During Run~2 of the LHC (2015--2018), the LHCb detector collected data with an instantaneous luminosity of $\mathcal{L}=4\times 10^{32}\cm^{-2}\sec^{-1}$.
In Run~3 (2022--2026), the instantaneous luminosity has been raised by a factor of five to $\mathcal{L}=2\times 10^{33}\cm^{-2}\sec^{-1}$. To face the increased particle rate, the LHCb detector underwent major hardware upgrades, termed ``Upgrade~I''. A new trigger entirely implemented in software has also been developed to reconstruct events at the LHC bunch crossing rate of $40\mhz$.

\begin{figure}[tb]
\begin{center}
\includegraphics[width=.95\textwidth]{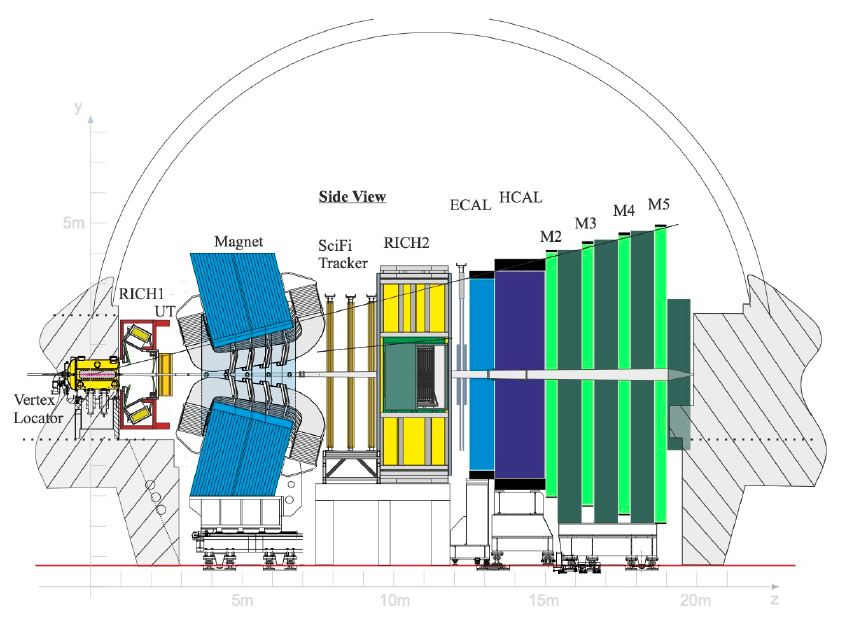}
\end{center}
\caption{The LHCb Upgrade~I detector~\cite{LHCb-DP-2022-002}. The muon stations in the terminal part of the spectrometer are depicted in light green, while the iron absorbers are indicated in dark green.}
\label{fig:lhcbdet}
\end{figure}

The LHCb Upgrade I detector~\cite{LHCb-DP-2022-002} is a general-purpose forward spectrometer covering the pseudorapidity range $2<\eta<5$, optimised for the detection of particles containing charm and beauty quarks. The detector is shown in Fig.~\ref{fig:lhcbdet} and includes a tracking system made of a silicon pixel detector called Vertex Locator (VELO), a silicon-strip detector called Upstream Tracker (UT) in front of the dipole magnet, and three Scintillating Fibre tracker (SciFi Tracker) stations downstream of the magnet. The particle identification is provided by two Ring Imaging Cherenkov detectors (RICH1 and RICH2) using C$_4$F$_{10}$ and CF$_4$ gases as radiators respectively, a shashlik-type Electromagnetic Calorimeter (ECAL), an iron-scintillator tile sampling Hadronic Calorimeter (HCAL), and four stations of Muon chambers (M2–M5) interleaved with iron absorbers.

The muon detector~\cite{LHCb-DP-2008-001} operated successfully during Run~1 (2010--2012) and Run~2 (2015--2018), delivering a single hit efficiency above 99\% per station with no sign of ageing~\cite{LHCb-DP-2012-002,LHCb-DP-2013-001,Albicocco:2019jvz} Several hardware interventions on the muon system were performed for the Upgrade~I to cope with the high luminosity~\cite{LHCb-TDR-014} and are briefly summarised in Sec.~\ref{sec:detector}. The muon detector operation during Run~3 is summarised in Sec.~\ref{sec:operation}, together with a novel method to estimate the luminosity using the muon chambers. The precise time and spatial calibrations are described in Sec.~\ref{sec:calibration}.

Restoring or possibly improving the Run 2 performance within a high background environment is the main challenge of the muon identification in the trigger. This requires to keep a high muon hit efficiency and to improve the rejection of misidentified hadrons due to combinatorial hits. The methods and new algorithms developed for this purpose in Run 3 are presented in Sec.~\ref{sec:performance}, where the performance is estimated by means of data calibration samples collected during the year 2024.

\section{The Upgrade~I muon detector}
\label{sec:detector}

\begin{figure}[tb]
\begin{center}
\includegraphics[width=.6\textwidth]{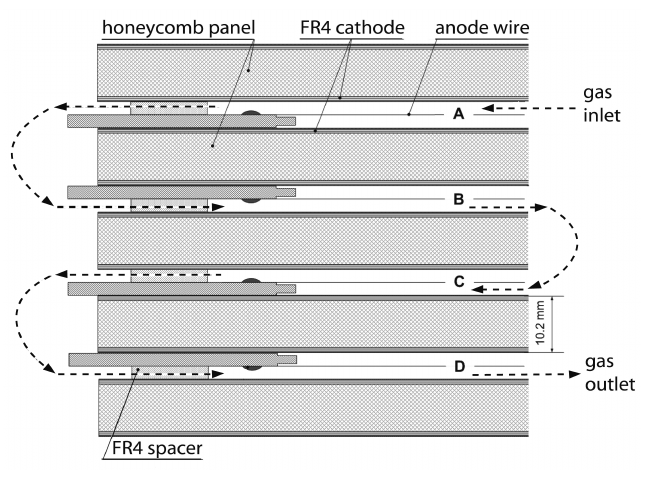}
\end{center}
\caption{Cross-section of a MWPC composed of four equal gas gaps of $5\mm$ thickness each with an anode wire plane in the
middle (referred to as A, B, C and D). The high-voltage is supplied to each gap separately.}
\label{fig:gaps}
\end{figure}

The Upgrade~I muon detector~\cite{LHCb-DP-2022-002} is composed of four rectangular stations M2--M5 placed sequentially along the beam line $(z)$ and interleaved with $80$-cm thick iron absorbers to filter penetrating tracks, as depicted in Fig.~\ref{fig:lhcbdet}. The stations are equipped with a total of 1104 Multiwire Proportional Chambers (MWPCs) covering a total area of $385~\rm{m}^2$. The MWPCs are made of four independent gas gaps (Fig.~\ref{fig:gaps}), each consisting of anode wires enclosed between two cathode planes, which are OR-ed to ensure redundancy and a nominal efficiency above 99\% within the $25\ns$ bunch spacing~\cite{LHCb-DP-2012-002}.

\begin{figure}[tb]
\begin{center}
\includegraphics[width=.50\textwidth]{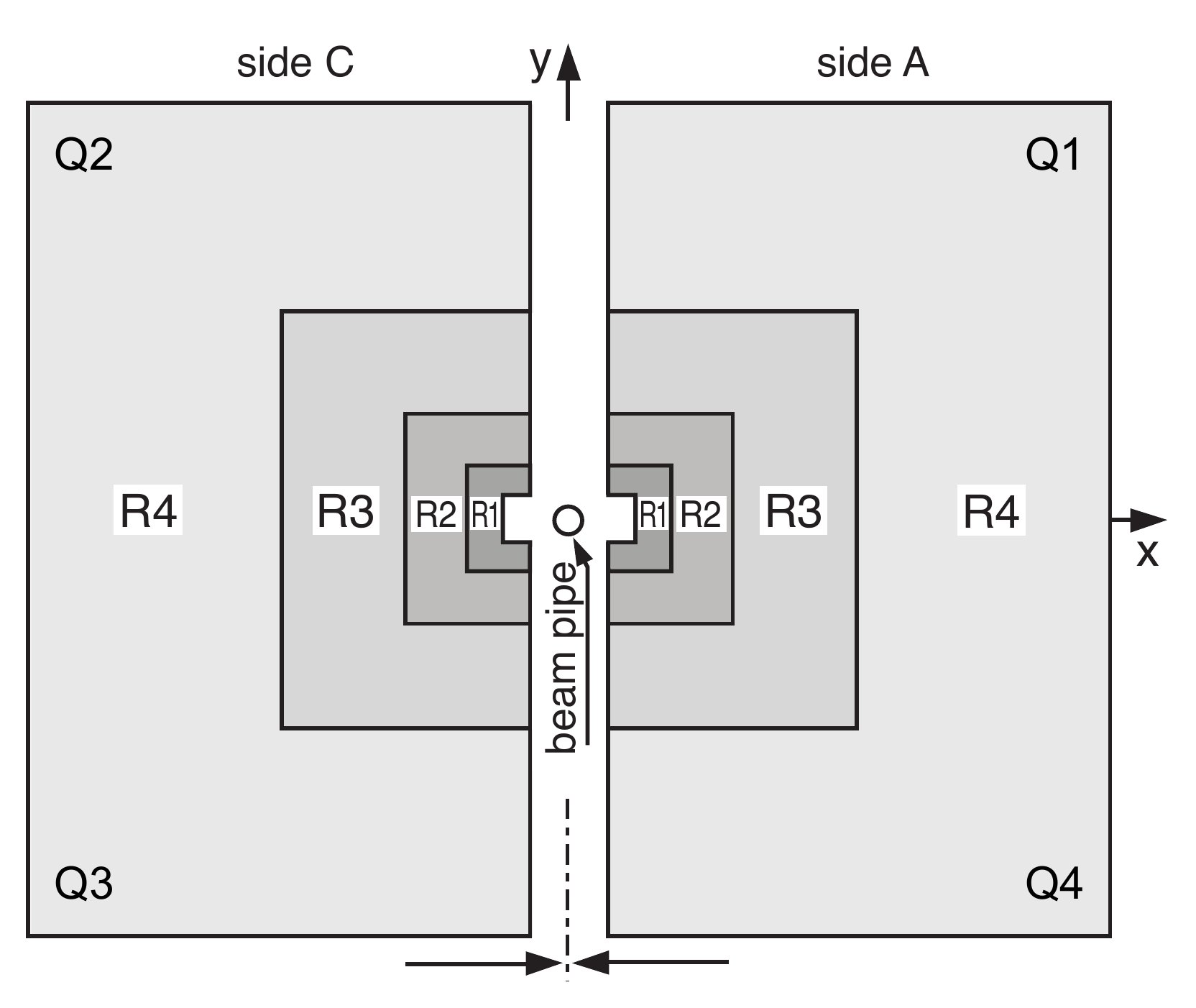}
\includegraphics[width=.46\textwidth]{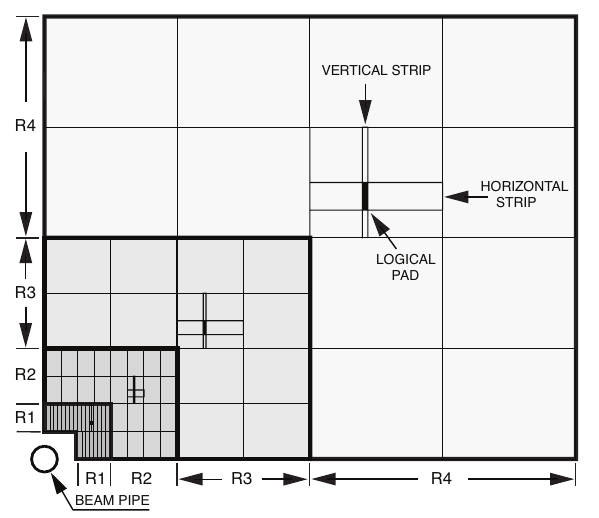}
\end{center}
\caption{Left: front view of a station with its side and region segmentation. The four quadrants Q1--Q4 are indicated in the R4 region. Right: front view of the Q1 quadrant of station M3, where the coincidence of horizontal and vertical strips (cathodic pads and group of wires, respectively) forms a logical pad~\cite{LHCb-TDR-004}.}
\label{fig:muondet}
\end{figure}

The stations are made of two mechanically independent and openable halves (side A and side C). Each station is divided into four quadrants (Q1--Q4) from the top right and counter-clockwise. Each quadrant is segmented into four regions (R1--R4) of decreasing readout granularity from the beam pipe to ensure uniform channel occupancy, as shown in Fig.~\ref{fig:muondet}. 

Although the muon system delivered nominal efficiency during Run~1 and Run~2, several hardware improvements were necessary to cope with the increased Run~3 luminosity. First, additional shielding elements were installed around the beam pipe section downstream of the calorimeters to reduce the rate of low-energy particles. In particular, a tungsten shielding was installed in place of the innermost HCAL cells, and two new beam-plugs were placed in correspondence of the HCAL and the M2 station, the latter being an improved version of the beam-plug used in Run~1 and Run~2~\cite{LHCb-TDR-014}. With this setup, a reduction of about $25\%$ of the particle rate is achieved in the innermost region (R1) of the M2 station. Second, the readout granularity was increased in regions R2, R3 and R4 of station M2 and in region R4 of station M5 by removing a logical layer responsible for grouping pads together during Run~1 and Run~2~\cite{LHCb-DP-2022-002}. Both these changes were aimed at reducing the expected inefficiency induced by the dead-time of the front-end electronics under high particle rates~\cite{Anderlini:2016hxw}.

The MWPCs operated at nominal efficiency during Run~1 and Run~2 with no appreciable signs of ageing. Most of the High-Voltage (HV) trips observed were recovered in-situ and found to be likely due to impurities produced during the construction process~\cite{Albicocco:2019jvz}. The chambers are expected to be fully operational for the whole lifetime of the Upgrade~I LHCb detector (ending in 2033) and, for the chambers in the external regions R3 and R4, to be further used during the high-luminosity Upgrade~II phase~\cite{ftdr,u2scoping} (ending in 2041). To ensure smooth operations, a number of spare chambers has been built and is available for replacements during the yearly technical stops of the LHC.

The Run~1 Front End Boards (FEBs) were kept since they were designed to sustain up to $100\kGy$ of radiation dose, which is sufficient to operate until the end of Run~4 (2033)~\cite{LHCb-DP-2022-002}. The FEBs are plugged onto the MWPCs and host an amplifier-shaper-discriminator stage implemented in a dedicated ASIC as well as a digital section that allows time alignment of the signals and grouping of physical channels. Depending on the detector region, the physical channels are either readout directly or combined into vertical and horizontal strips and processed by the new Off-Detector Electronics (nODE), which allows the full event readout at $40\mhz$ and represent the main upgrade of the readout electronics. Each nODE board is equipped with four radiation-tolerant custom ASICs, known as nSYNC~\cite{Cadeddu:2019uzq}. These ASICs perform clock synchronization, bunch crossing alignment, data hit production, and time measurements. The nODE employs four optical unidirectional uplinks, which use GigaBit Transceiver chipset (GBTx) and Versatile Link components~\cite{Moreira:2009pem} to transmit formatted and tagged data to the back-end electronics. The back-end electronics is made of custom PCIe40 boards, configured via the firmware for data acquisition (TELL40 boards) or control (SOL40 boards)~\cite{LHCb-DP-2022-002}. At this stage, logical pads are defined by the coincidence of horizontal and vertical strips, as depicted in Fig.~\ref{fig:muondet}. The event builder servers host the TELL40 boards together with graphic processing units running the first level of the new fully-software trigger named High Level Trigger 1 (\texttt{HLT1}). This first stage employs a few tens of inclusive trigger lines in which the real-time reconstruction and selection are performed at $30\mhz$, representing the average rate of visible interactions. The \texttt{HLT1} output rate of around $1\mhz$ is fed into the second trigger stage (\texttt{HLT2}) in which hundreds of lines perform a more refined reconstruction and the exclusive selection of physics channels running on a farm of general purpose CPU servers~\cite{Aaij:2019uij}. 
Both trigger levels employ a real-time muon identification as described in Sec.~\ref{sec:performance}.

\section{Detector operation}
\label{sec:operation}

To deliver high efficiency and to keep the rate of HV trips at a minimum, dedicated hardware settings were introduced in the early phase of Run~3 following the detector response to the luminosity increase.

The MWPCs are filled with an Ar/CO$_2$/CF$_4$ gas mixture with 38:57:5 volume proportions. The gas system works in closed mode with a recirculation efficiency above 95\%. The gas flow in each MWPC varies between $1.0$ and $1.5\litre/\!\hr$, depending on the station and region. This corresponds to an input of $87\litre/\!\hr$ of fresh mixture which includes $6\litre/\!\hr$ of consumption for gas analysis and ensures more than one volume change per day.

Although all MWPCs have 5-mm-thick gas gaps, different HV values are set in each station and region, since the chambers have different pad size and readout technique~\cite{LHCb-DP-2012-002}. The HV values lie within the working range $2.53-2.62\kvolt$, corresponding to a gas gain in the range $(4-7)\times10^4$~\cite{Dane:2007hs}. For safety reasons, the HV is lowered to $2.0\kvolt$ at the end of each LHC fill and set to its nominal value a few minutes before collisions occur. In the case of HV trip, the gaps are excluded from data-taking and recovered in situ, similar to the procedure adopted during Run~1 and Run~2~\cite{Albicocco:2019jvz}. Since four independent HV lines are used for the gaps in each chamber, these failures do not affect the detector efficiency. Five chambers were replaced following an irreversible damage due to electrical discharge since the beginning of Run~3.

The FEB thresholds were set to 7 times the equivalent noise charge (ENC)~\cite{Sarti:1131816} to deliver nominal efficiency while keeping the electronic noise below $1\khz$ per channel. At this level, the noise contribution to the trigger rate is negligible~\cite{LHCb-DP-2012-002}. A noise scan on the whole detector is repeated once or twice a year, while noisy channels are monitored and tuned during data-taking.

With the above settings, stable conditions are achieved while operating at the chamber efficiency plateau~\cite{LHCb-DP-2008-001}, as confirmed by the hit efficiency measurement described in Sec.~\ref{ssec:hit}.

At the start of the physics data-taking at nominal luminosity, nODE-TELL40 desynchronisations were observed at a rate of about 14 links per hour, leading to DAQ inefficiencies of the order of 1\%. This rate was reduced to 2 links per hour by raising the $7\volt$ power supply of the GBTx chips to $8\volt$, which improved the load regulation of the DC-DC converters feeding the GBTx. The rate of occurrence was further reduced by improving the TELL40 firmware error handling and the data decoding in the software trigger. As the same issue was observed by other subdetectors, a centralised procedure was implemented to automatically resynchronise the affected readout system, reducing the DAQ inefficiency from this source to a negligible level. 
During the year 2024, the total LHCb DAQ efficiency was about 95\%, with the muon system contributing with 0.3\% to the overall downtime.

Online monitoring of hit maps, hit multiplicity, and other reconstructed observables was also implemented and is available for control room operators via the web-based application \texttt{MONET}~\cite{Adinolfi:2017ntr}.

\subsection{Integrated charge}
\label{ssec:age}

An estimation of the maximum integrated charge is performed to assess the MWPC ageing expected at the end of Run~4.

The maximum current across all gaps in a given station and region is recorded at nominal luminosity. To estimate the maximum integrated charge $Q^{\rm{max}}_{\rm{int}}$ for a given chamber type, the current value is multiplied by the number of days during which the LHC has delivered collisions, weighted by the average instantaneous luminosity. In Table~\ref{tab:1}, the maximum integrated charge per centimetre of wire is provided for the period 2022--2025, together with the expected value for the whole Run~3--Run~4 period (2022--2033).

\input{tab1.tex}

As expected from the particle flux, the integrated charge decreases rapidly from the inner region, closer to the beam pipe, towards the outer regions. The flux also decreases from M2 to M4 due to the shielding of the iron filters, with M5 receiving additional flux from backsplash events~\cite{Paluch:1630392}.
At the end of Run~2, a maximum accumulated charge of $630\mC/\!\cm$ was estimated in the chambers belonging to the most irradiated region (M1R2)~\cite{Albicocco:2019jvz}, with no sign of an efficiency drop. At the end of Run~4, the value expected in M2R1 exceeds this tested level: however, should any degradation in performance be observed, a number of spare chambers is already available for replacement. Conversely, the low expected charge in the chambers belonging to the outer regions R3 and R4 allows for their operation also for the high-luminosity Upgrade~II phase~\cite{ftdr,u2scoping}.

\subsection{Online luminosity monitor from chamber currents}
\label{ssec:lumi}

The luminosity at the LHCb interaction point is levelled to $2\times10^{33}\cm^{-2}\sec^{-1}$ by adjusting the transverse displacement of the LHC beams~\cite{Follin:2014nva}. In Run~3, a dedicated system named Probe for LUminosity MEasurement (PLUME) has been installed to monitor the machine luminosity by means of an array of quartz radiators~\cite{LHCb-TDR-022,LHCB-FIGURE-2025-008}. Measurements from various subdetectors allow to independently validate the luminosity measurement and serve as beam monitors in case of PLUME unavailability. Protection against unwanted luminosity fluctuations is in fact especially important for the detector elements close to the beam pipe. To this end, a novel method has been developed to monitor the luminosity based on MWPC currents.

The currents drawn by the MWPCs are proportional to the intensity of the incoming radiation, which is proportional to the machine luminosity. Since the currents are asynchronously monitored by the HV distribution modules~\cite{LHCb-TDR-004}, they can be used to monitor the luminosity online independently of the DAQ system. Two effects are considered and corrected for in this measurement. The first correction accounts for the gain variation with the gas pressure. Since the temperature in the LHCb cavern is kept constant, the current increase with the atmospheric pressure is approximately linear~\cite{Dane:2007hs}. This correction is usually below 1\%. A second correction is necessary to compensate for the additional particle flux generated by the beam-gas collisions in the System for Measuring Overlap with Gas (SMOG2) storage cell~\cite{BoenteGarcia:2024kba}.
The SMOG2 cell is filled with gases to collect fixed-target collisions at LHCb in parallel with the nominal $\proton\proton$ data-taking. When the gas is injected in SMOG2, the chamber currents increase by a few percent, corresponding to the additional rate of particles produced in beam-gas collisions. This effect is subtracted from the current readings whenever a gas injection is performed.

\begin{figure}[tb]
    \centering
    \includegraphics[width=0.8\textwidth]{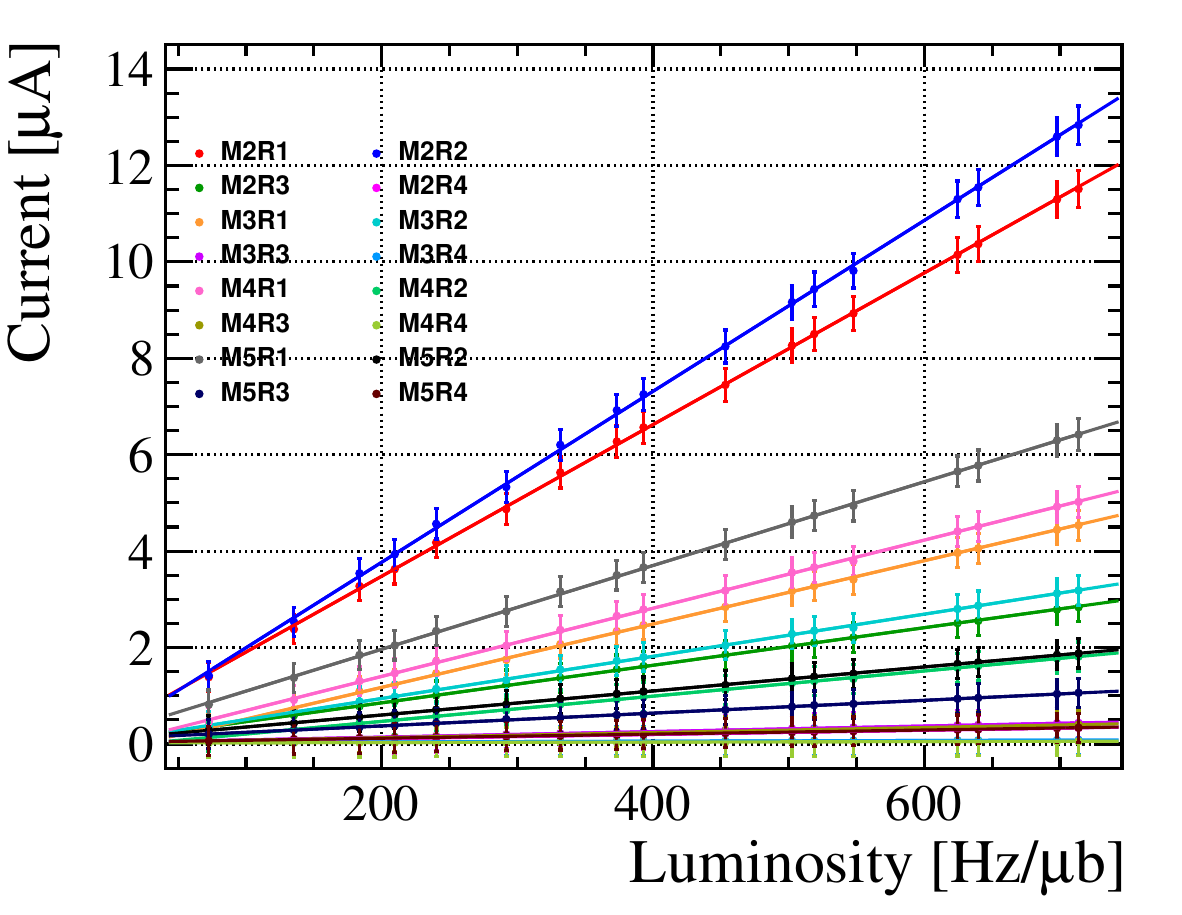}
    \caption{Average currents for each station and region of the detector as a function of the instantaneous luminosity. Linear fit functions are overlaid.}
    \label{fig:muscan}
\end{figure}

After the above corrections, the current values are converted into luminosity values by means of a calibration point provided by PLUME. Fig.~\ref{fig:muscan} shows the average current values across MWPC gaps of each station and region of the detector as a function of the luminosity measured by PLUME. These data were collected in May 2025 during a special run in which the luminosity was changed in small steps within a short period of time.

The luminosity estimate is performed once per second and has a precision of a about 7\%, mainly determined by the statistical fluctuations of the current readings. This precision is enough to detect anomalous luminosity values, providing an independent safety measure against luminosity spikes.

\section{Detector calibration}
\label{sec:calibration}
The time alignment of the readout channels is mandatory to reach nominal detection efficiency within the 25-ns-spaced bunch crossings. A first time alignment was performed after the installation of the new readout electronics and is continuously monitored online, as explained in Sec.~\ref{ssec:time}. The spatial alignment of the muon stations is needed to ensure the proper hit positions, which are exploited in the muon identification. This is checked whenever the muon stations are opened and closed for maintenance, as reported in Sec.~\ref{ssec:space}.

\subsection{Time alignment}
\label{ssec:time}
A precise time alignment of the nODEs readout channels is paramount to ensure high hit reconstruction efficiency. Signal arrival times can differ across channels, for example due to analog processing delays and cable length differences. Time delays can be adjusted both at the readout channel level (nODE) and at the physical channel level (FEB). Coarse delays can be set in the nODEs by modifying dedicated registers with values from 0 to 7 where each step corresponds to a $25\ns$ shift. Similarly, fine delays in FEB channels can take values from 0 to 31, where each step corresponds to a $1.6\ns$ shift.

\begin{figure}[tb]
\begin{center}
    \includegraphics[width=0.49\textwidth]{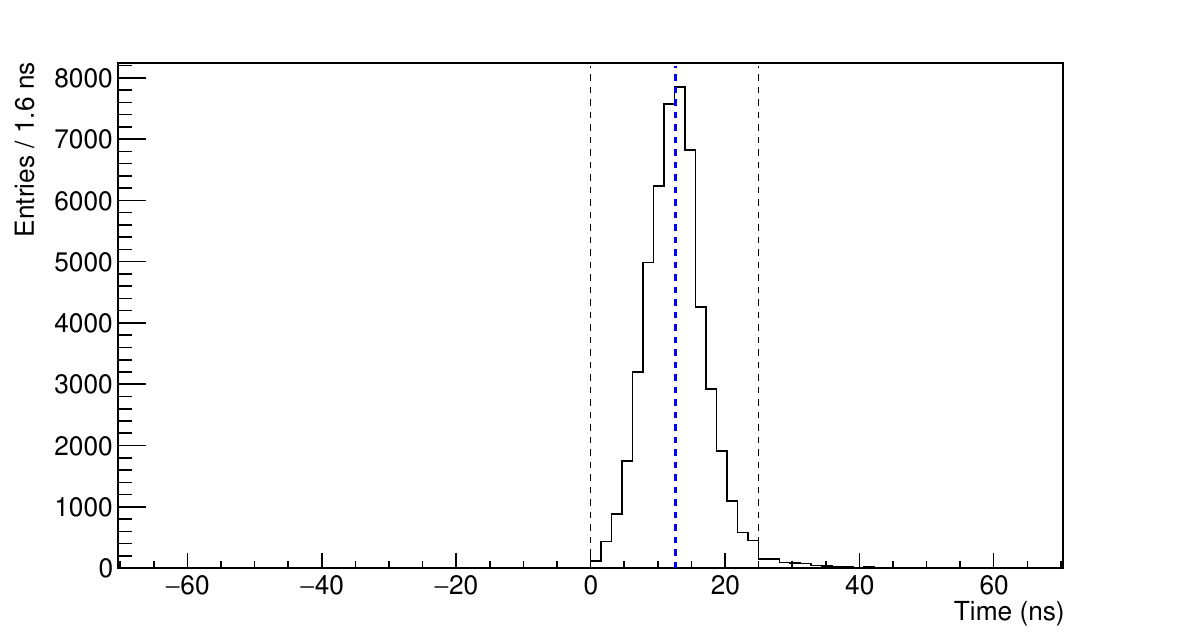}
    \includegraphics[width=0.49\textwidth]{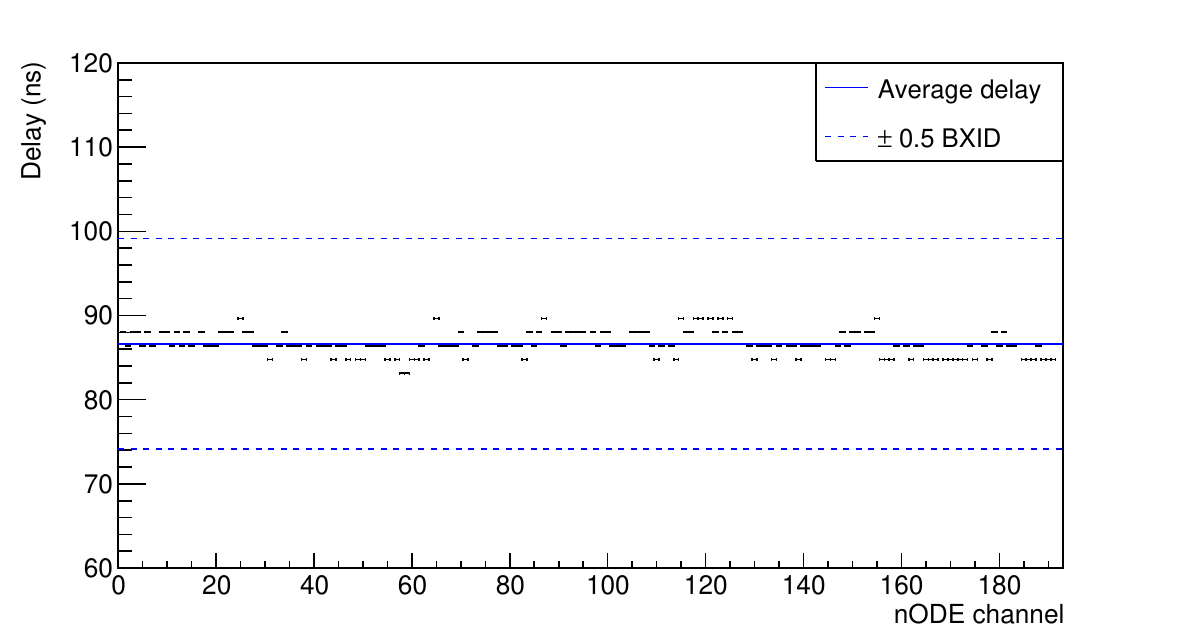}
\end{center}
    \caption{Left: time spectrum for a nODE channel in region R3 of station M4. The blue vertical bars represent the reference BXID to align to. Right: Time delays for each channel in one nODE belonging to station M2.}
    \label{fig:time}
\end{figure}

In Run~3, the time alignment procedure has been completely revised with respect to the one used in Run~1 and Run~2, profiting from the increased instantaneous luminosity. The higher rate of muons in fact allows to align nODE channels individually, whereas neighbouring channels were previously grouped and aligned together~\cite{LHCb-DP-2012-002}. To perform the time alignment, special collision data named Time Alignment Events are required. These data are collected by triggering on isolated colliding proton bunches, \ie those in which neighbouring bunches are empty, so that muon hits can be unambiguously assigned to a given Bunch Cross Identifier (BXID). The data are recorded with a larger time window of $125\ns$ in order to reconstruct events from hits arriving with a delay exceeding the nominal time window of $25\ns$.
Only hits belonging to a muon track segment are considered by requiring at least 3 spatially aligned hits across at least three stations. An example time distribution for one nODE channel is shown in Fig.~\ref{fig:time} (left). In this typical time spectrum, the muon signal is contained in the $25\ns$ window, with a longer tail towards higher times due to drift time and time walk, delayed cross-talk hits and longer path of low momentum tracks~\cite{LHCb-DP-2012-002}.

The time correction is determined by the difference between the median time and the BXID centre, and applied such that both the rising edge and part of the right tail are included in the BXID. Fig.~\ref{fig:time} (right) shows the delays applied to each channel of one nODE, where variations of a few nanoseconds are expected due to differences in cable lengths. In case of noisy channels, in which a muon peak is not visible, the delays of the neighbouring channel with the same cable length are applied. These corrections can be checked a-posteriori by verifying that neighbouring channels which are independently aligned have similar delay values.

After a first complete time alignment, nominal hit efficiency values were measured on data, as described in Sec.~\ref{ssec:hit}. The time alignment is repeated once or twice a year or whenever a chamber is replaced, with typical adjustments comparable to the time resolution of $1.6\ns$.

\subsection{Spatial alignment}
\label{ssec:space}

After each year of data-taking, the muon stations are usually opened to perform regular maintenance. After closing, the position of each station is measured via optical references with an accuracy of a few millimeters. A more precise alignment measurement is subsequently performed by matching muon track segments to well reconstructed tracks.

A dedicated trigger line is used to select pairs of muons with an invariant mass compatible with that of the $J/\psi$ meson, and originating from a vertex which is detached with respect to the primary $\proton\proton$ collision vertex. These requirements significantly reduce the combinatorial background and enrich the sample with $B \to J/\psi X$ decays. Only ``long tracks'' are considered, \ie those reconstructed from hits across all tracking detectors (VELO, UT and SciFi), once their respective spatial alignment is completed. Muon segments are defined by clusters of hits in at least three different stations that align with a straight-line trajectory pointing to the interaction point.

The alignment procedure employs an iterative Kalman filter that minimizes the total $\chi^2$ of a set of tracks while adjusting the $x$ and $y$ positions of the muon sides A and C, as detailed in Ref.~\cite{LHCb-DP-2012-002}. Adding other degrees of freedom, such as the positions of the individual chambers, would require much larger data samples and is subject to systematic uncertainties due to the distribution of the tracks and the large size of the logical pads. 

At the beginning of 2024, once all the tracking detectors were aligned, misalignments in the muon detectors exceeding 1~mm from the nominal position were observed in the $x$ direction, with displacements up to approximately 10~mm. In the $y$ direction, all stations were found to be aligned within 1~mm. These displacements are corrected for by offsetting the detector positions in the dedicated conditions database~\cite{LHCb-DP-2022-002,Valassi:2013sna}.

\section{Performance}
\label{sec:performance}

The real-time muon identification is fundamental for the LHCb physics program, as a sizeable fraction of the physics analyses contains muons in the final state. The goal of the Upgrade~I muon detector was to keep the Run~1--Run~2 identification performance after a fivefold increase in the instantaneous luminosity~\cite{LHCb-TDR-014}. This goal requires the muon identification algorithms to significantly reduce the misidentification probability, mainly due to the increased amount of combinatorial hits, with a small impact on the muon efficiency. 

This goal is realised in two steps: a loose boolean selection called \texttt{IsMuon}~\cite{LHCb-DP-2013-001}, developed in Run~1, and a $\chi^2_{corr}$ algorithm, specifically developed for Run~3~\cite{LHCb-DP-2020-002}. \texttt{IsMuon} requires the presence of hits in consecutive muon stations within a Field Of Interest (FOI) and depending on the track momentum, as reported in Table~\ref{tab:2}. The FOI is centred around the track extrapolation to the muon system and its size is parametrized as
\begin{equation}
    \text{FOI} = a + b\times \exp{(-cp)},
    \label{eq:foi}
\end{equation}
where $p$ is the track momentum and the parameters $a$, $b$ and $c$ depend on the station and region and have been determined from simulation~\cite{LHCb-DP-2013-001,Lanfranchi:1202759}. In Run~3, the size of the FOI has been reduced by 20\% to reduce the misidentification probability, especially at low momentum, with a loss of muon identification efficiency below 1\%.

\input{tab2.tex}

The $\chi^2_{corr}$ is computed on the hits passing the \texttt{IsMuon} selection and accounts for their correlation due to the multiple scattering experienced by a muon traversing the main absorbers of the LHCb detector, namely ECAL $(25~\rm{X_0})$, HCAL $(53~\rm{X_0})$ and the muon iron filters ($47.5~\rm{X_0}$ each), shown in Fig.~\ref{fig:lhcbdet}.
This is defined as:
\begin{equation}\label{eq:chi2corr}
\chi^2_{corr} = \delta\overrightarrow{x}^T \text{V}_x^{-1} \delta\overrightarrow{x} + \delta\overrightarrow{y}^T \text{V}_y^{-1} \delta\overrightarrow{y},
\end{equation}
where $\delta\overrightarrow{x}$ and $\delta\overrightarrow{y}$ are the horizontal and vertical distances between the track extrapolation points and the position of the closest muon hit for each station. The covariance matrices $\text{V}_x$ and $\text{V}_y$ both have a diagonal contribution from the detector resolution and off-diagonal terms quantifying the correlations induced by the multiple scattering, as detailed in Ref.~\cite{LHCb-DP-2020-002}. 

The \texttt{IsMuon} and $\chi^2_{corr}$ algorithms are implemented in both trigger levels \texttt{HLT1} and \texttt{HLT2}, where in the former case the looser \texttt{IsMuon} selection is commonly used. At \texttt{HLT2}, the $\chi^2_{corr}$ is converted into a likelihood by integrating and then subtracting the $\chi^2_{corr}$ distributions from simulated samples of muons (signal) and protons (background). The muon likelihood is then combined with that of the RICH and calorimeters to build a global variable with improved particle identification performance.

Two performance analyses based on 2024 data are presented. Sec.~\ref{ssec:hit} reports the efficiency of muon hit detection for each region and station, while Sec.~\ref{ssec:muid} shows the muon identification performance using large calibration samples of muons and charged hadrons.

\subsection{Hit efficiency} 
\label{ssec:hit}

To measure the hit efficiency, detached $J/\psi \to \mu^+\mu^-$ decays are selected with a dedicated trigger line. The background contamination is further reduced by requiring one track in the final state to be positively identified as a muon (the \textit{tag} track) while no muon information is used on the other one (the \textit{probe} track), which can therefore be used to measure the efficiency. Both tracks are required to be within the muon geometrical acceptance and to have a momentum greater than $10\gevc$ in order to be able to penetrate all three iron absorbers~\cite{Lanfranchi:1202759}. The coordinates of a track traversing the muon system are obtained by linearly extrapolating the trajectory provided by the tracking system. The extrapolated coordinates are provided with negligible uncertainty with respect to the spatial resolution of the muon detector.

\begin{figure}[tb]
    \centering
    \includegraphics[width=0.7\textwidth]{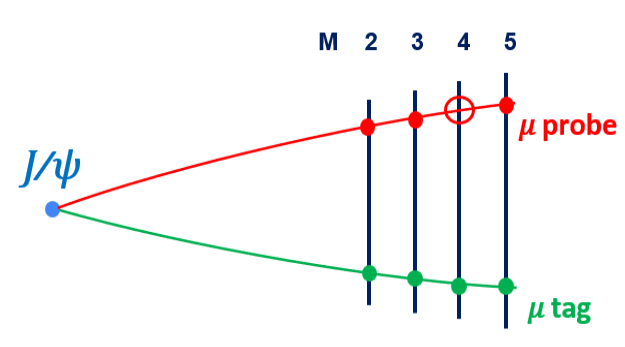}
    \caption{Sketch of the method used to measure the hit efficiency of station M4. Details are provided in the main text.}
    \label{fig:scheme}
\end{figure}

As depicted in Fig.~\ref{fig:scheme}, to measure the efficiency of a station (open red circle), the presence of hits associated to the track extrapolation is required in the remaining three stations (full red circles). The sample purity is further enhanced by requiring the tag track to have associated hits in all four muon stations (full green circles).

\begin{figure}[tb]
    \centering
    \includegraphics[width=0.9\textwidth]{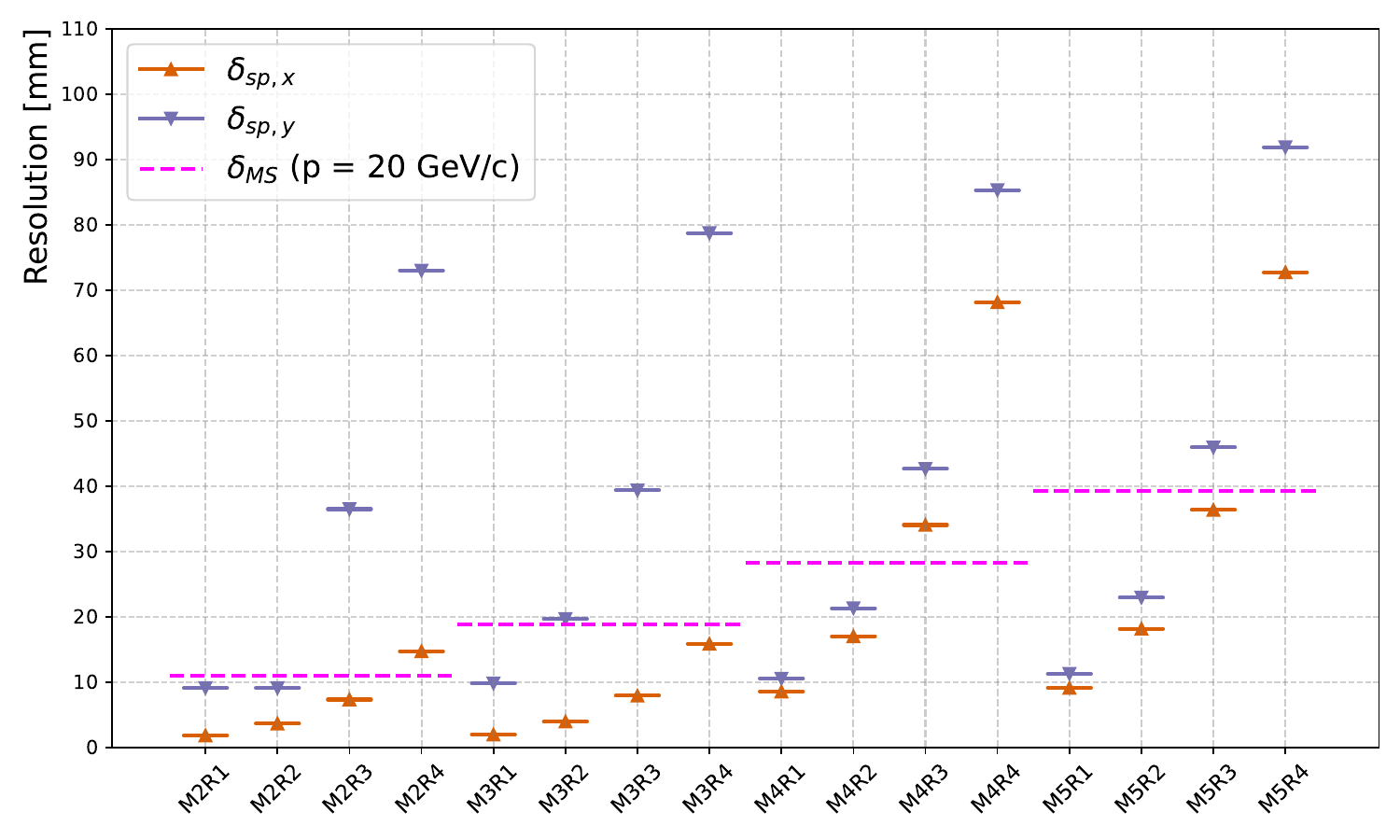}
    \caption{Horizontal (red) and vertical (blue) spatial resolutions of each station and region of the detector. The average multiple scattering deviation for a track with $p=20\gevc$ is shown in magenta.}
    \label{fig:res}
\end{figure}

The association of a hit to the extrapolated track ($\rm{ex}$) is based on their relative horizontal and vertical distance, divided by the uncertainty:
\begin{equation}
    \sigma_x = \frac{x_{\rm{hit}}-x_{\rm{ex}}}{\delta_x} 
    \leq n
    ~~~ \& ~~~
    \sigma_y = \frac{y_{\rm{hit}}-y_{\rm{ex}}}{\delta_y}
    \leq n.
    \label{eq:sigma}
\end{equation}
The uncertainty $\delta_{x,y}$ receives contributions from the spatial resolution $(\rm{sp})$ of the chamber and from the average multiple scattering deviation $(\rm{MS})$ experienced by the track:

\begin{align}
    \delta_{(x,y)}^2 = \delta^2_{\rm{sp},(x,y)} + \delta^2_{\rm{MS}} =
    \left( \frac{\rm{pad}_{(x,y)}}{\sqrt{12}} \right) ^2 + \delta^2_{\rm{MS}}
    \label{eq:delta}
\end{align}

where $\rm{pad}_{x,y}$ is the horizontal or vertical size of a logical pad. The values of the spatial resolution are reported in Fig.~\ref{fig:res} for each region and station of the detector.

Similar in magnitude~\cite{LHCb-TDR-004}, the multiple scattering contribution $\delta_{\rm{MS}}$ on each station is parametrised from simulated data in which the full detector material traversed by a muon is accounted for. No significant difference is observed between the vertical and horizontal multiple scattering displacements, as expected from the detector geometry. The effect of the track polar angle is also found to be negligible. An example of the multiple scattering contribution to the resolution is shown in Fig.~\ref{fig:res} for a muon with a momentum of $20\gevc$. The distribution of $\sigma_{x,y}$ (Eq.~\ref{eq:sigma}) is found to be approximately Gaussian. 

To select good muon candidates, the four hits associated to the tag track are required to be as close as possible to its extrapolated coordinates. A distance of $n=2$ standard deviations is found to provide a high signal purity, whereas tighter thresholds reduce the signal yield significantly. 

A station is considered efficient if at least one hit is found within a searching window around the probe track. Different sizes for the searching window are tested. The efficiency is found to reach a plateau at around $n=4$, so that $n=5$ is chosen, after which the background contribution becomes significant. The residual background is in fact subtracted by computing the hit efficiency in a searching window opened in the opposite quadrant with respect to the probe track position, where only background hits are expected. With the chosen $n=5$ searching window, the background contribution to the efficiency is usually at the permille level. The hit multiplicity in the searching window is approximately described with a Poissonian distribution and varies significantly across the detector regions, with the probability of finding two or more hits varying between 20\% and 60\%. 

\begin{figure}[tb]
    \centering
    \includegraphics[width=0.9\textwidth]{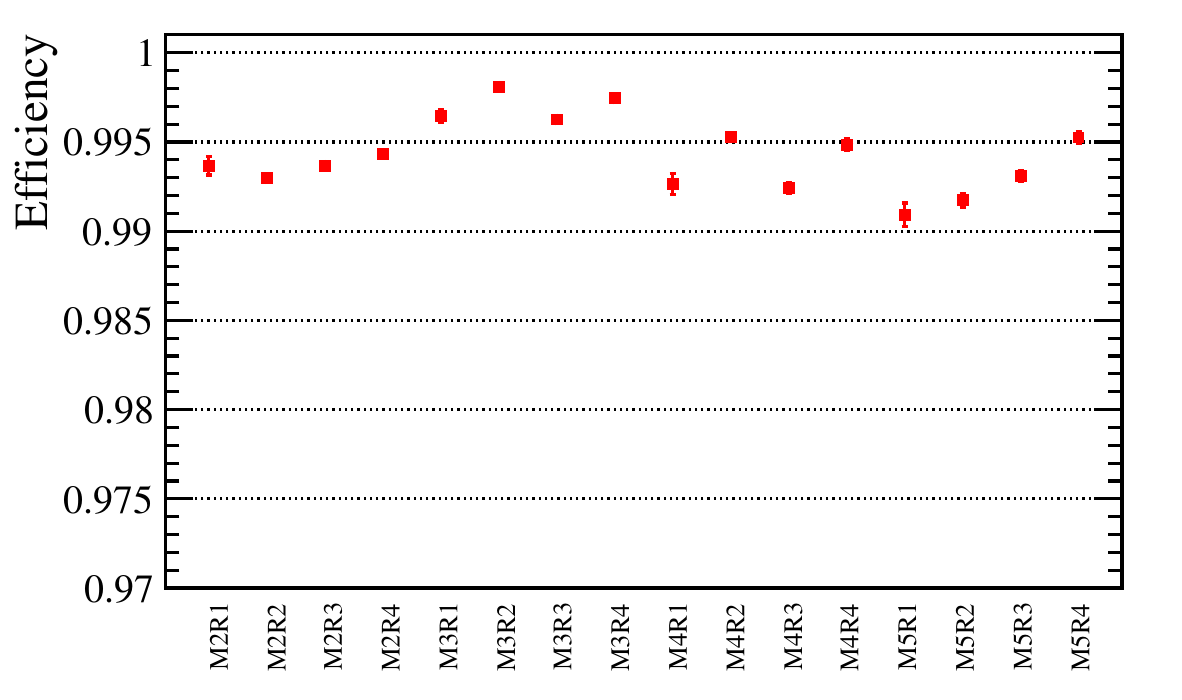}
    \caption{Hit efficiency for every station and region of the detector. The statistical uncertainties are shown but some are too small to be visible. A few malfunctioning chambers have been excluded in the measurement.}
    \label{fig:hiteff}
\end{figure}

Fig.~\ref{fig:hiteff} shows the hit efficiency measured on a small sample of proton collision recorded in 2024, corresponding to an integrated luminosity of $88\invnb$.

All regions are found to meet the 99\% efficiency requirement. Thanks to the high statistics and the simplicity of the method, an online monitor of hit efficiency was implemented in Run~3, enabling precise measurements with just a few hours of collision data. This allows to promptly identify and address inefficiencies during data-taking.

\subsection{Muon identification performance}
\label{ssec:muid}

In this section, the combined performance of the \texttt{IsMuon} and $\chi^2_{corr}$ variables are provided, representing the muon identification performance achieved by exploiting the pattern of hits in the muon system. 

For this study, calibration data collected during October 2024 are used, corresponding to an integrated luminosity of $1.2\invfb$. A large sample of muons is provided by detached $J/\psi \to \mu^+\mu^-$ decays, while pions, kaons, and protons are obtained from $K^0_S\to \pi^+\pi^-$, $\phi(1020) \to K^+K^-$ and $\Lambda \to p\pi^-$ decays, respectively. These samples can be reconstructed with high purity without using particle identification on the probe track, which can therefore be used to determine the muon identification efficiency and the hadron misidentification probabilities. The probability to misidentify a charged hadron as a muon depends in fact on the amount of random muon hits with an additional component of true muons from decays-in-flight in the case of pions and kaons, which is especially relevant at low momentum. 

The events are required to be selected by the trigger independently of the presence of the probe track, in order to remove the correlation between the trigger selection and the muon detector response. The tracks in the final state are required to be within the muon detector geometrical acceptance, to have a track-fit $\chi^2$ per degree of freedom smaller than 3 and to have a probability of being reconstructed as a fake track, \ie not corresponding to an actual particle trajectory, smaller than 30\% as determined by the multivariate operator \texttt{GhostProb}~\cite{DeCian:2255039}. The purity of the samples is further enhanced by requiring the tag track to pass stringent particle identification requirements. 

\begin{figure}[!ht]
    \centering
    \includegraphics[width=0.48\linewidth]{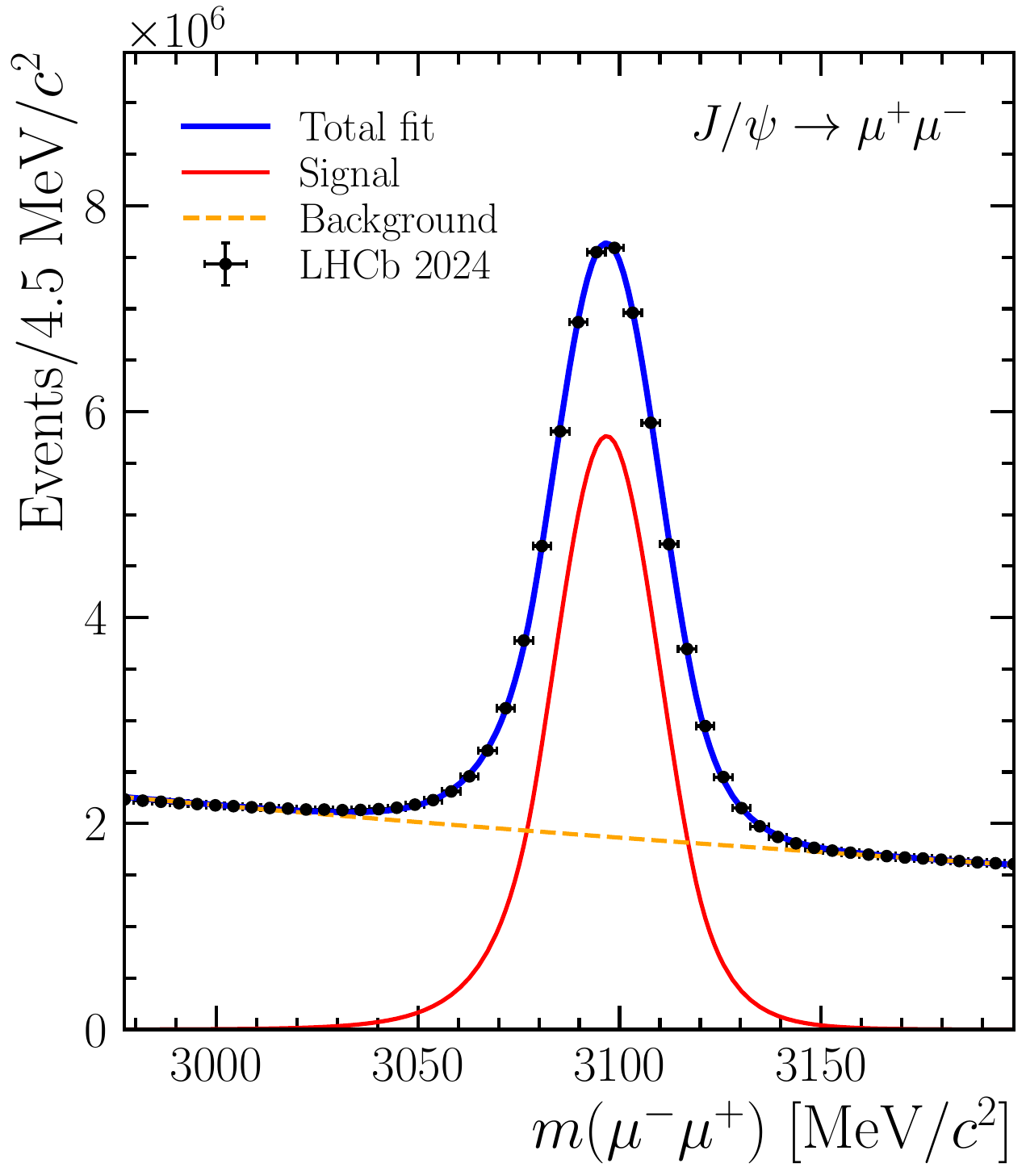}
    \includegraphics[width=0.50\linewidth]{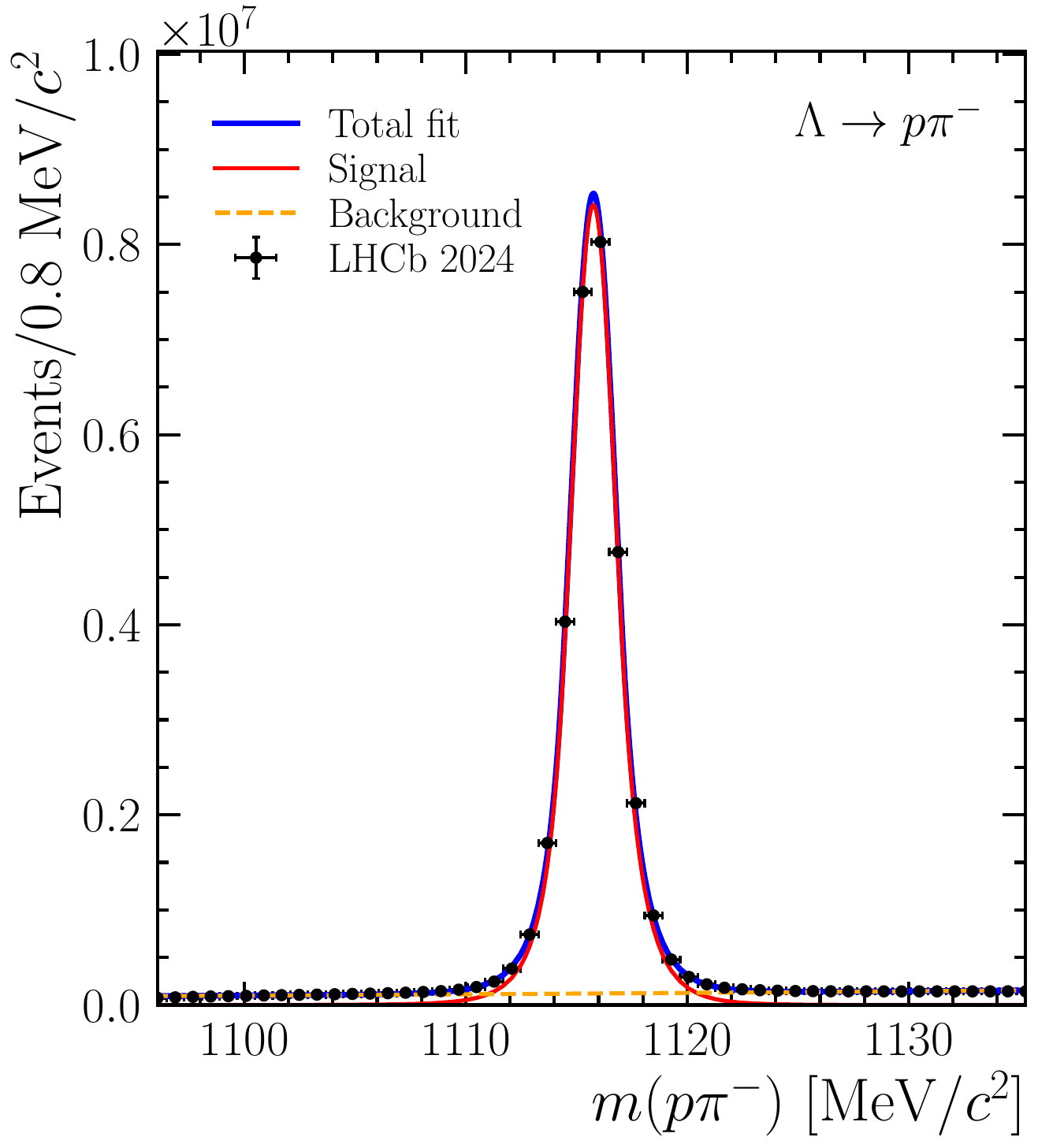}
    \includegraphics[width=0.48\linewidth]{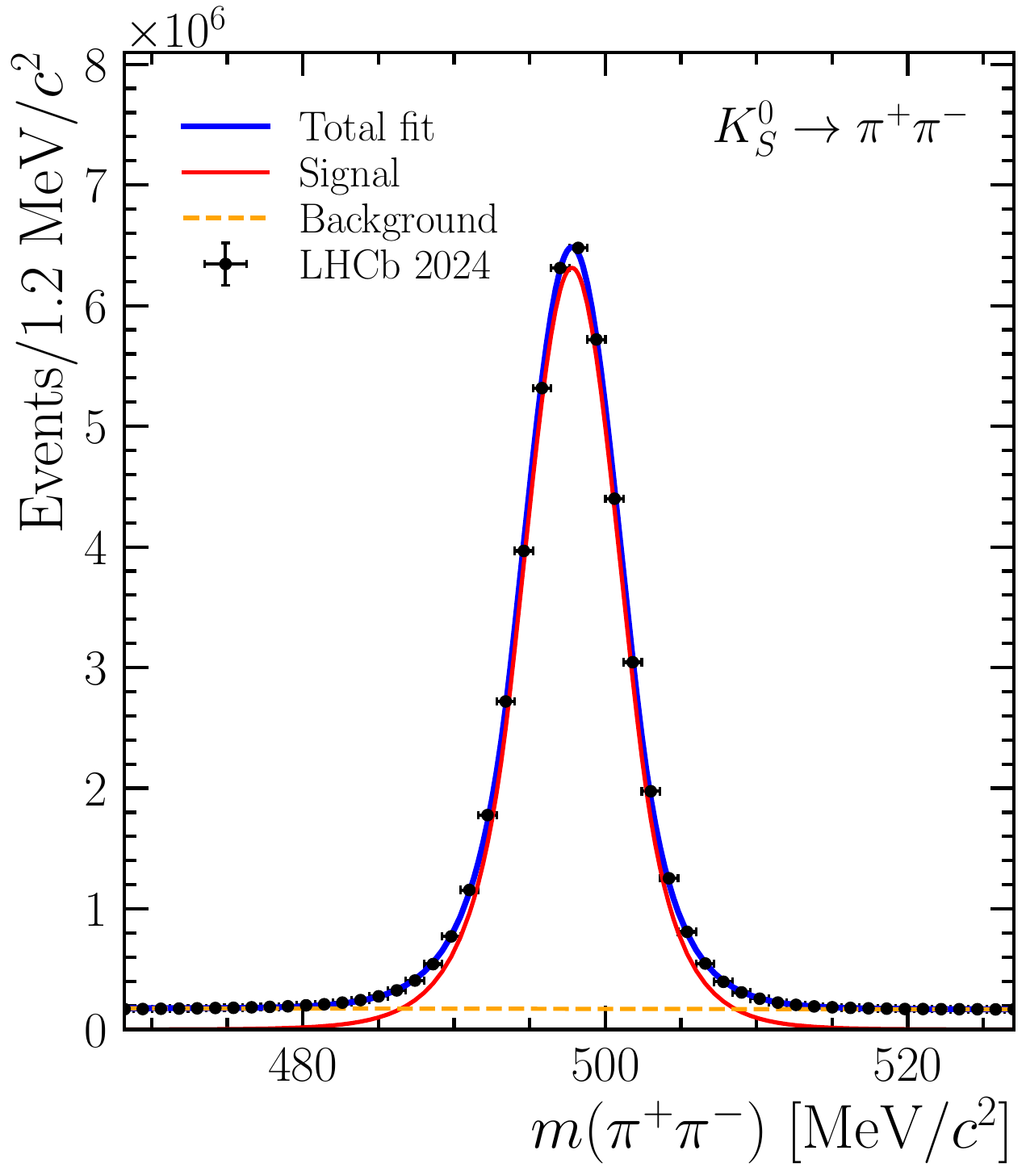}
    \includegraphics[width=0.50\linewidth]{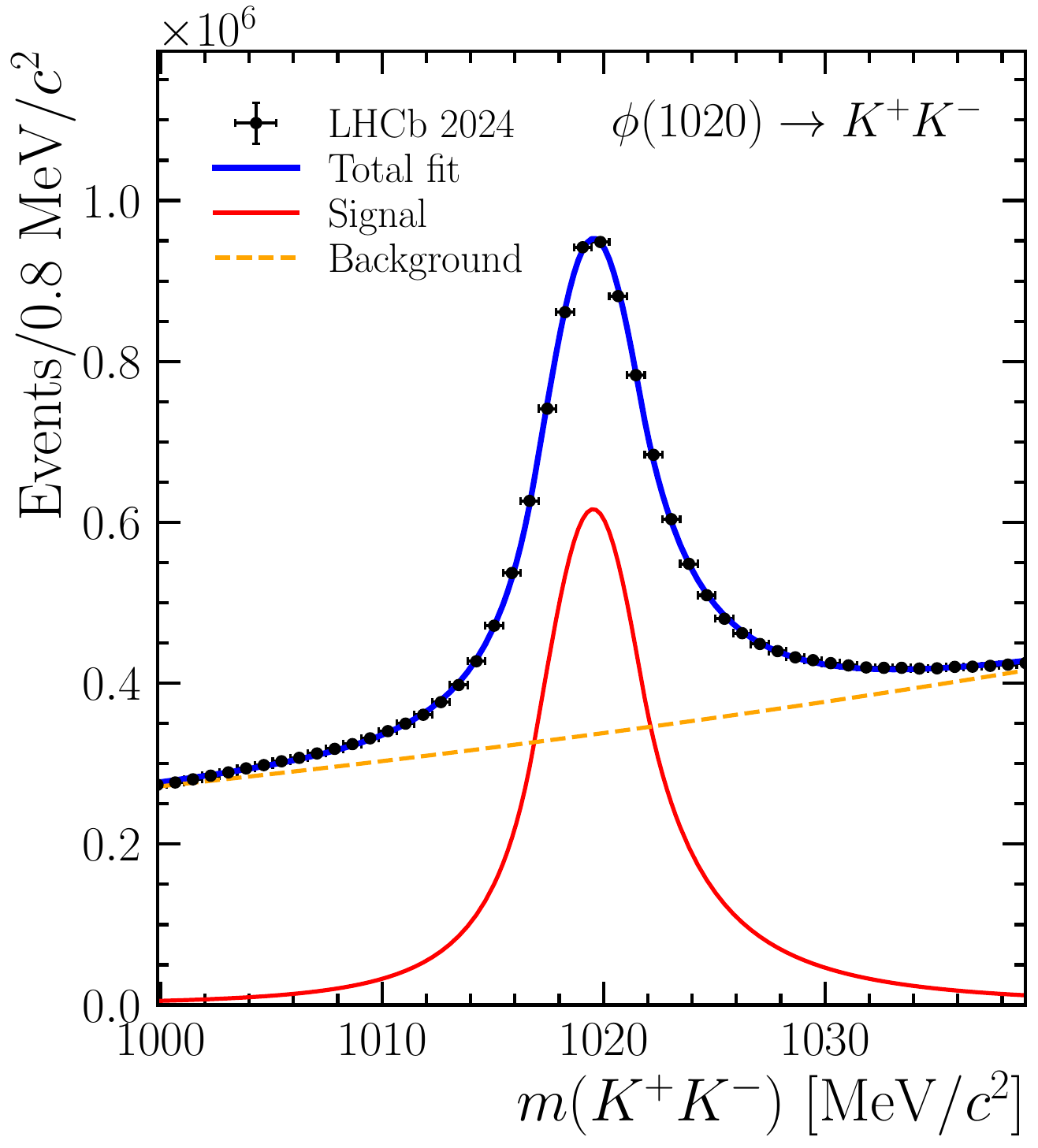}

    \caption{Invariant mass spectra of the calibration samples with the fit curves overlaid.}
    \label{fig:mass}
\end{figure}

\begin{figure}[!ht]
    \centering
    \includegraphics[width=0.7\linewidth]{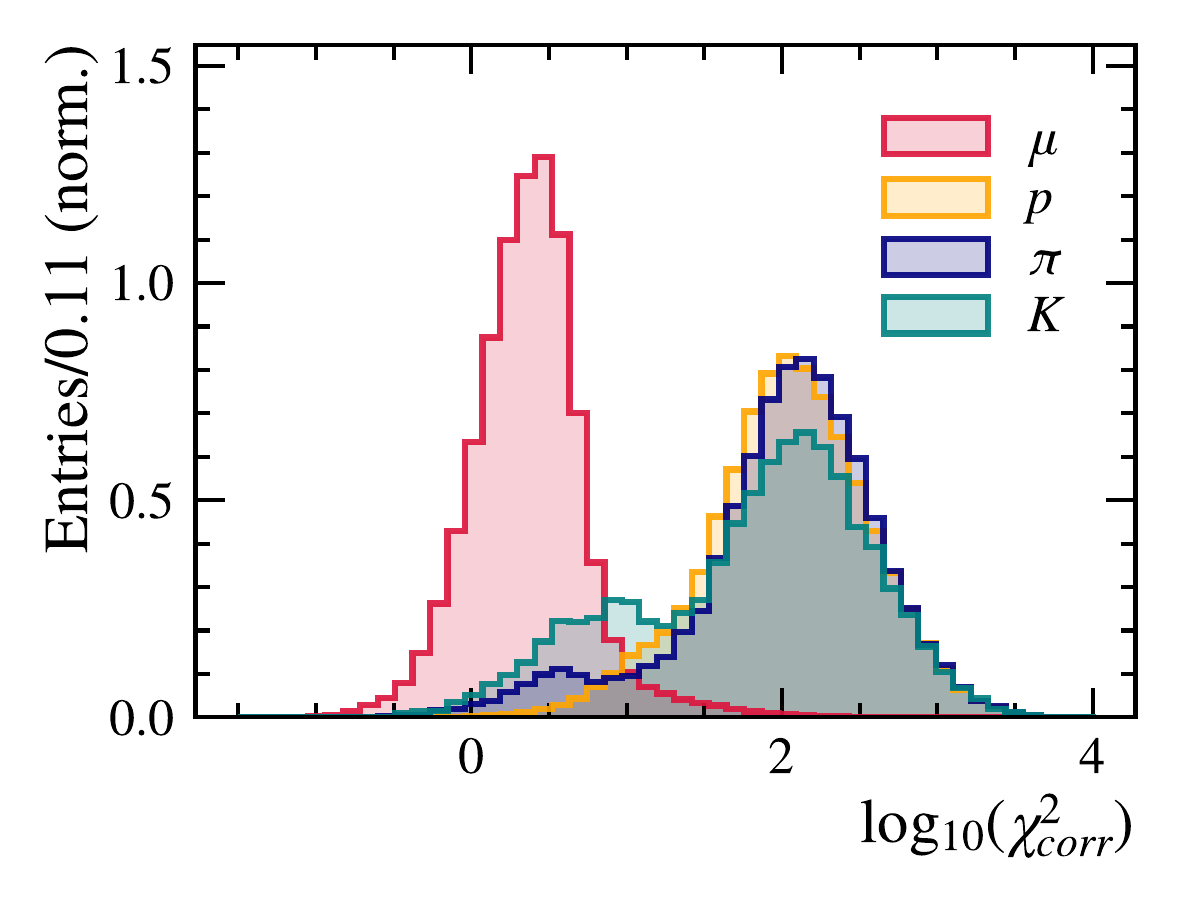}
    \caption{Background subtracted distributions of $\log(\chi^2_{corr})$ for muons and hadrons with a minimum momentum of $10\gevc$. The contribution of decays in flight is visible as a peak at lower $\chi^2_{corr}$ values in the pion and kaon distributions.}
    \label{fig:chi2}
\end{figure}

The residual combinatorial background in each data sample is statistically subtracted by means of the \sPlot technique~\cite{Pivk:2004ty}. First, a fit is performed on the invariant mass (\textit{discriminating variable}) where signal and combinatorial distribution are known. The fit model is composed of a double-tailed Crystal Ball function~\cite{Skwarnicki:1986xj} for the signal and an exponential function for the background, as shown in Fig.~\ref{fig:mass}. Then, from the fit covariance matrix, \sPlot provides per-event weights with which the signal and background contributions of the $\chi^2_{corr}$ (\textit{control variable}) can be unfolded. The resulting background-subtracted $\chi^2_{corr}$ distributions are shown in Fig.~\ref{fig:chi2} for each particle species. 

The efficiencies and misidentification probabilities of a given cut can finally be computed as the sum of weights for the events passing the cut divided by the total. A systematic uncertainty associated with the choice of the fit model is included in the following results. This is defined as the uncertainty required to account for the discrepancy between the efficiencies obtained with the nominal fit function and those obtained using an alternative fit function, such that the two estimates agree within one standard deviation. 

\begin{figure}[!ht]
    \centering
    \includegraphics[width=0.7\linewidth]{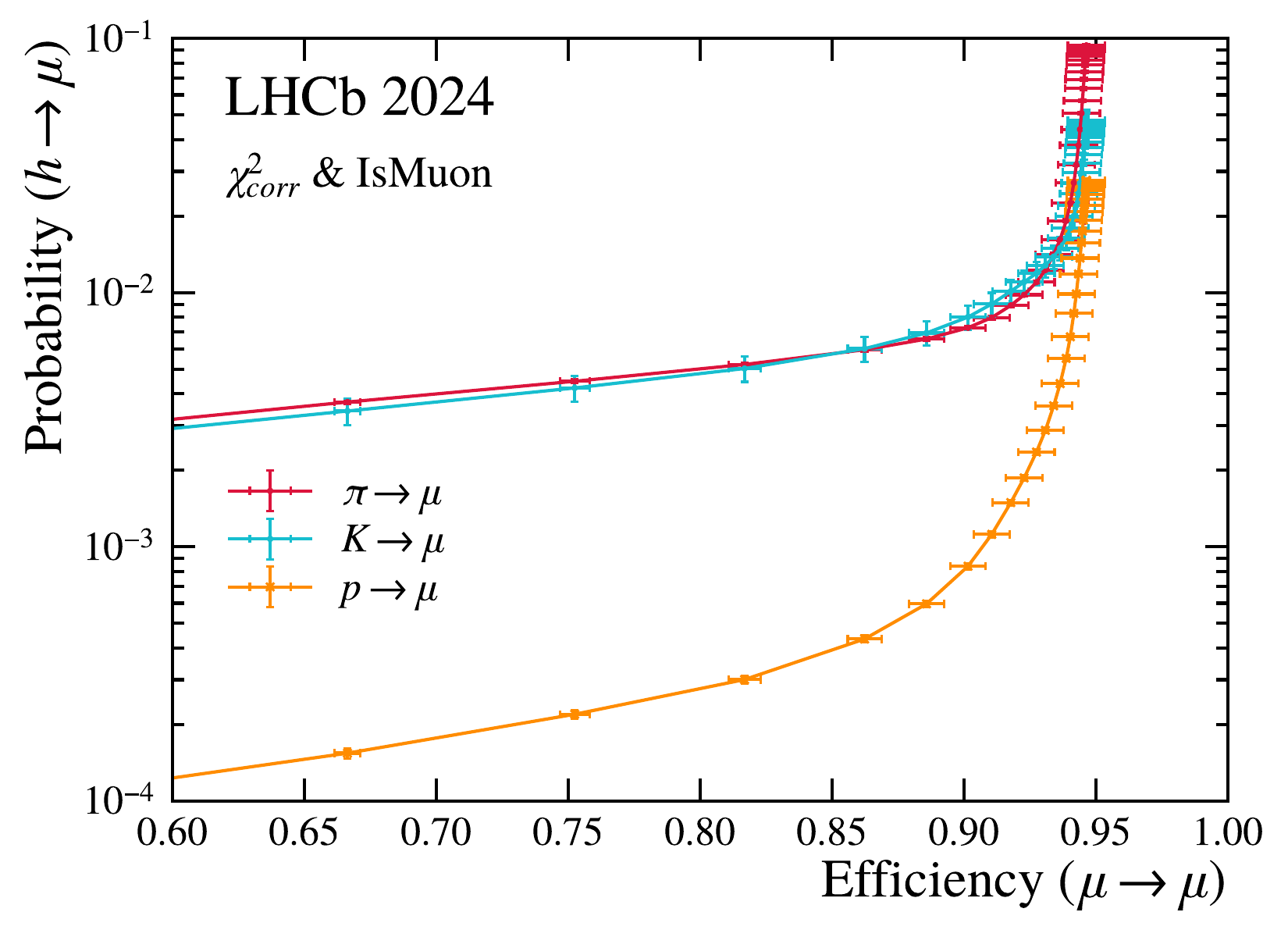}
    \caption{ROC curve of the product of the \texttt{IsMuon} and $\chi^2_{corr}$ operators for each hadron type for tracks having a momentum greater than 10 GeV/c and a transverse momentum greater than 0.8 GeV/c.}
    \label{fig:ROC}
\end{figure}

The muon efficiency versus hadron misidentification probability (ROC curve) for each hadron type is computed for various $\chi^2_{corr}$ cuts and shown in Fig.~\ref{fig:ROC}. All tracks are required to have a momentum greater than $10\gevc$ and a transverse momentum greater than $0.8\gevc$, which are common requirements for tracks selected in the analysis of charm and beauty hadron decays. For muon identification efficiencies above $90\%$, misidentification probability of a few permille is observed on pions and kaons, and an order of magnitude lower for protons.

\begin{figure}[!ht]
    \centering
    \includegraphics[width=0.49\linewidth]{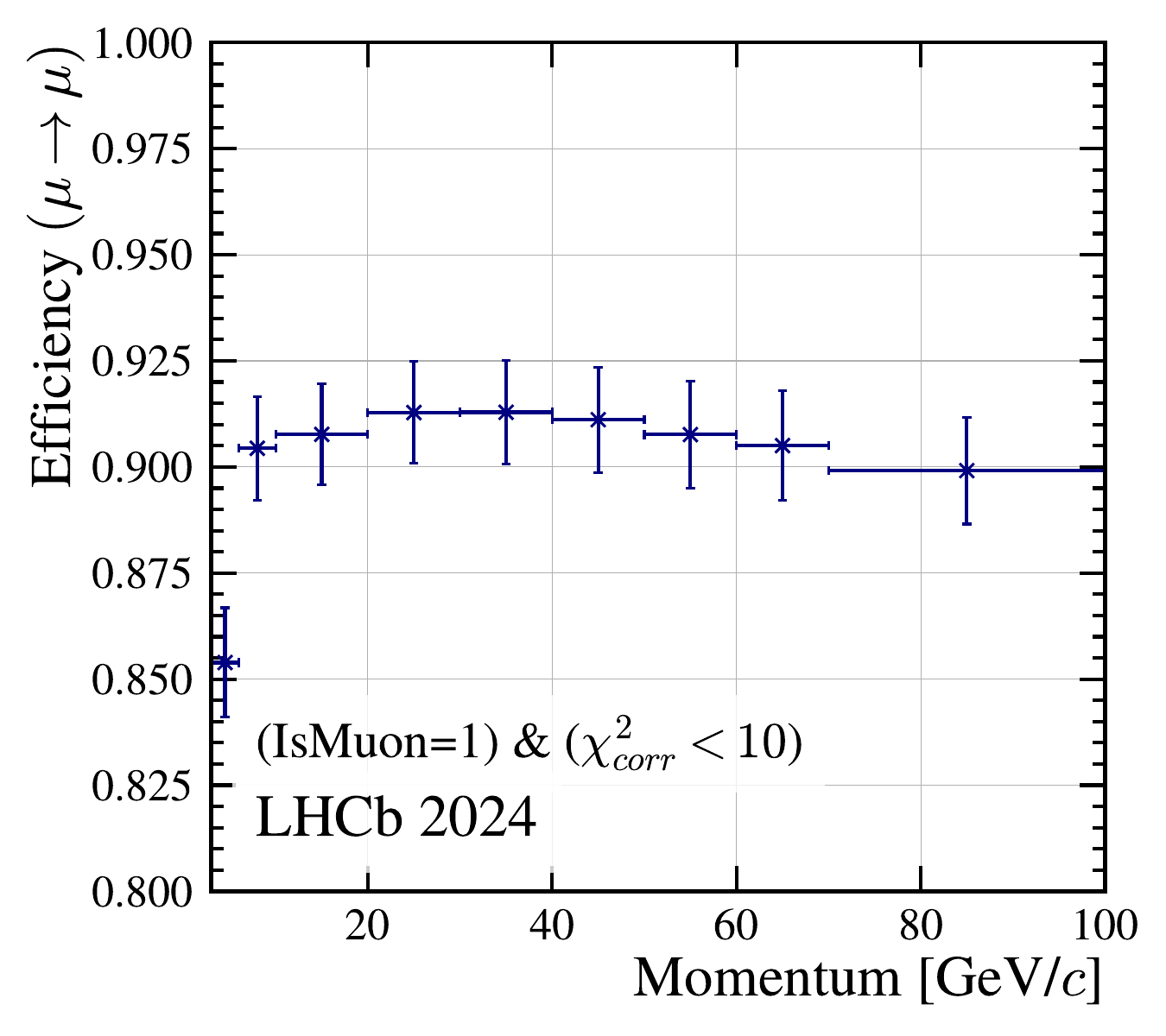}
    \includegraphics[width=0.49\linewidth]{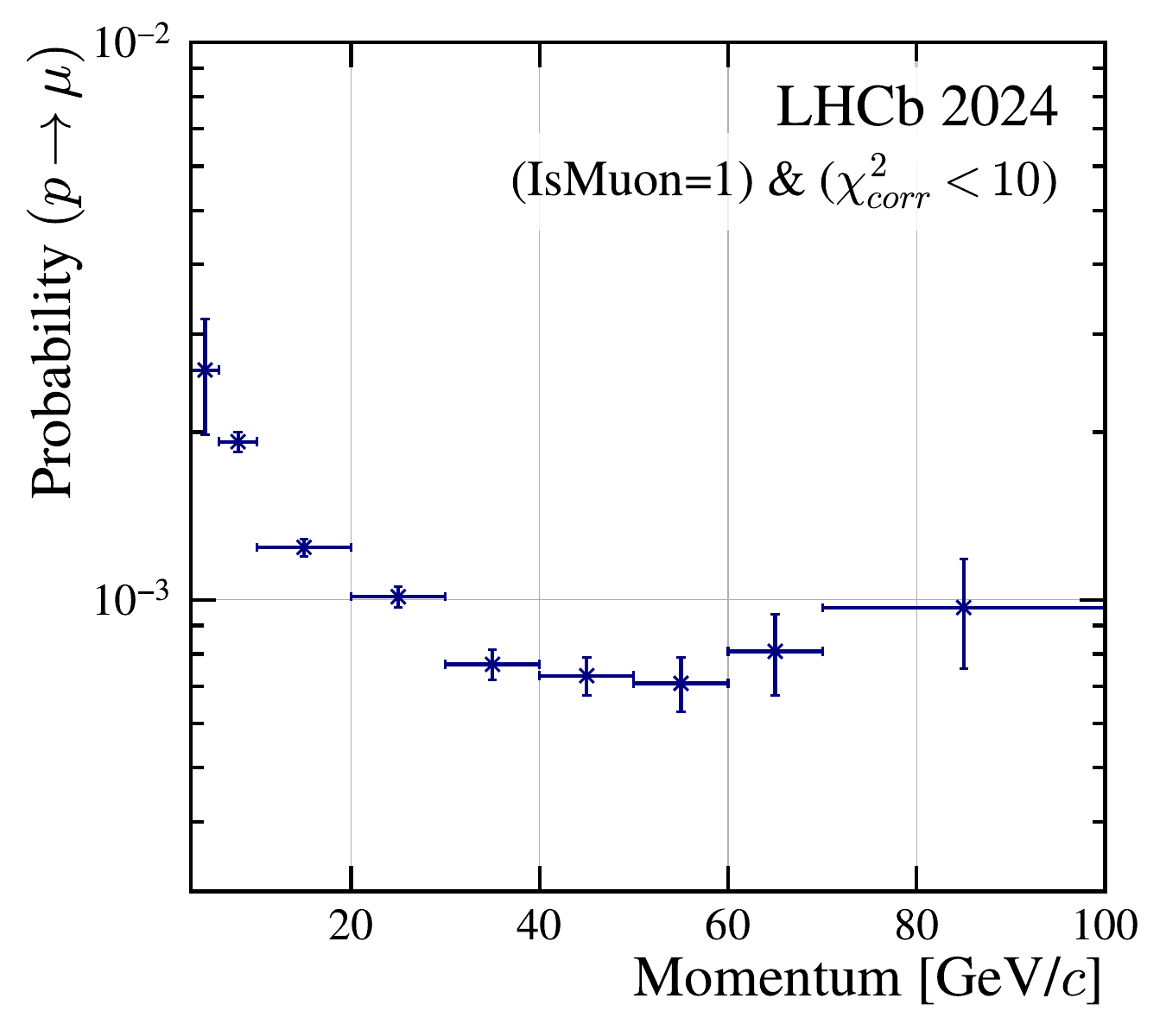}
    \includegraphics[width=0.49\linewidth]{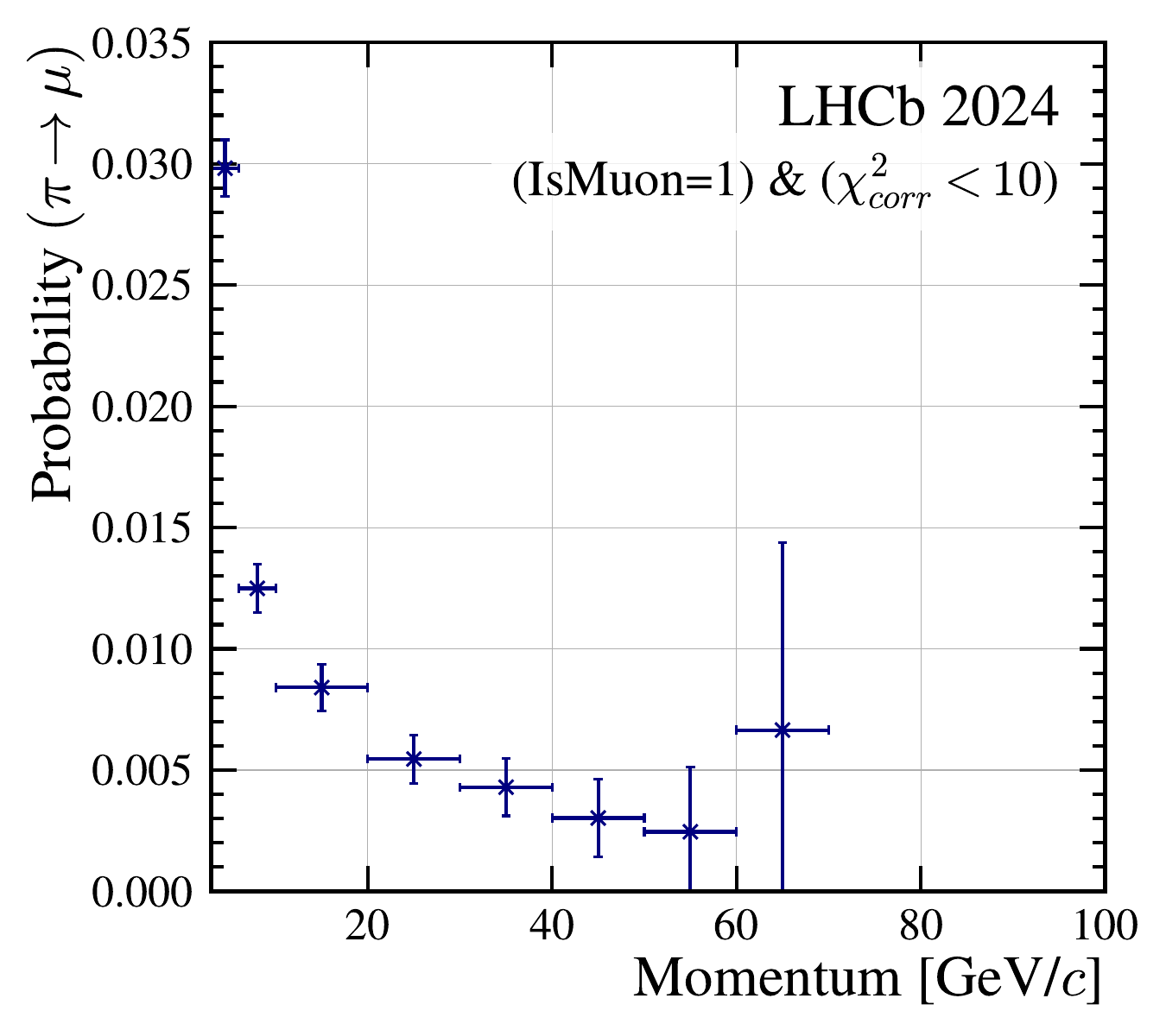}
    \includegraphics[width=0.49\linewidth]{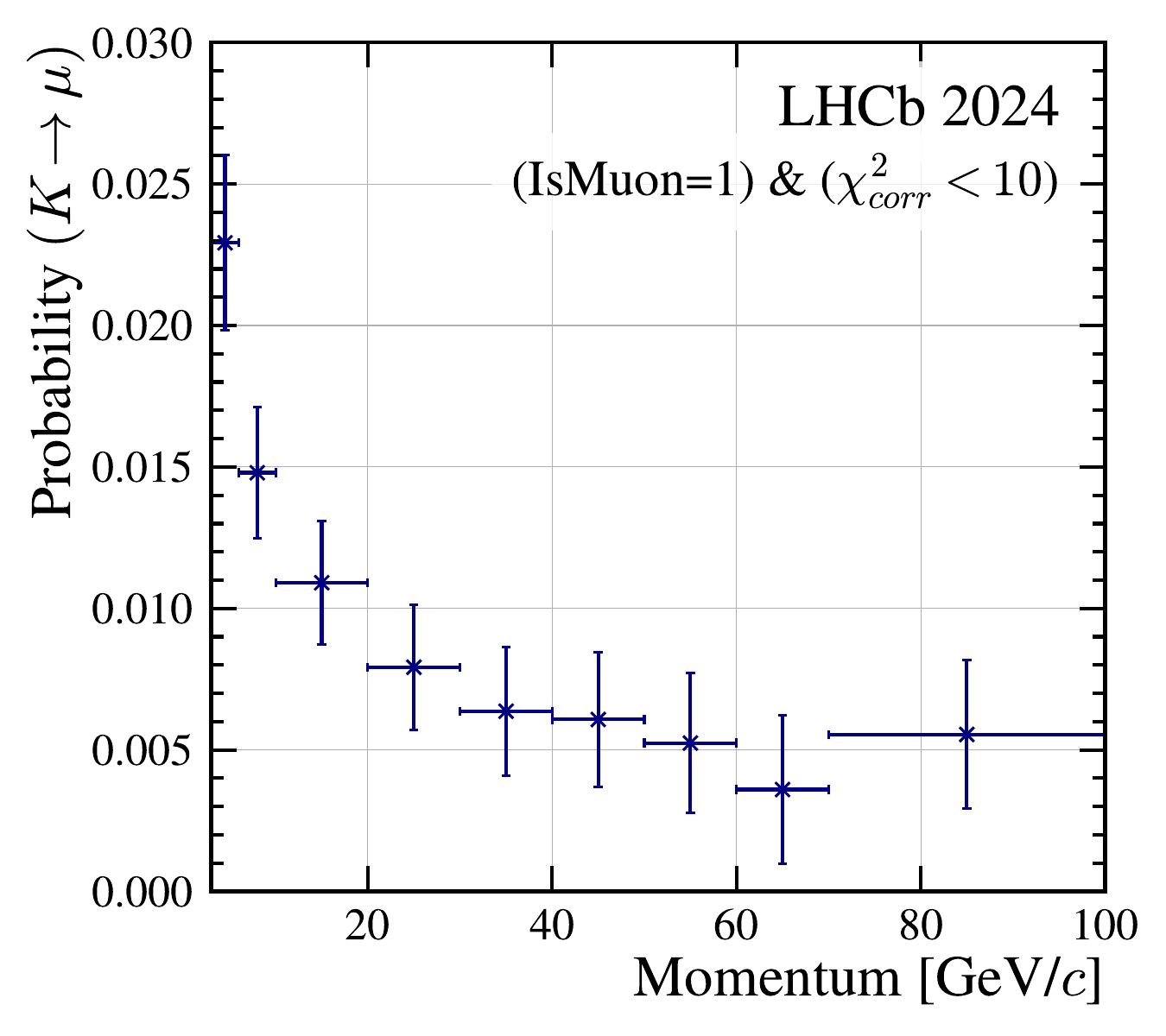}
    
    \caption{Efficiency and hadron misidentification probability for the \texttt{IsMuon} and $\chi^2_{corr}<10$ selection in different momentum ranges.}
    \label{fig:misid_pbins}
\end{figure}

As the muon identification depends on the track kinematics, the efficiencies and hadron misidentification probability are also provided in Fig.~\ref{fig:misid_pbins} as a function of the momentum for tracks with $\pt>0.8\gevc$, which shows that most of the misidentification probability is concentrated at low momentum, where the \texttt{IsMuon} FOI is larger and the hit resolution is dominated by the multiple scattering.

\begin{figure}[!ht]
    \centering
    \includegraphics[width=0.49\linewidth]{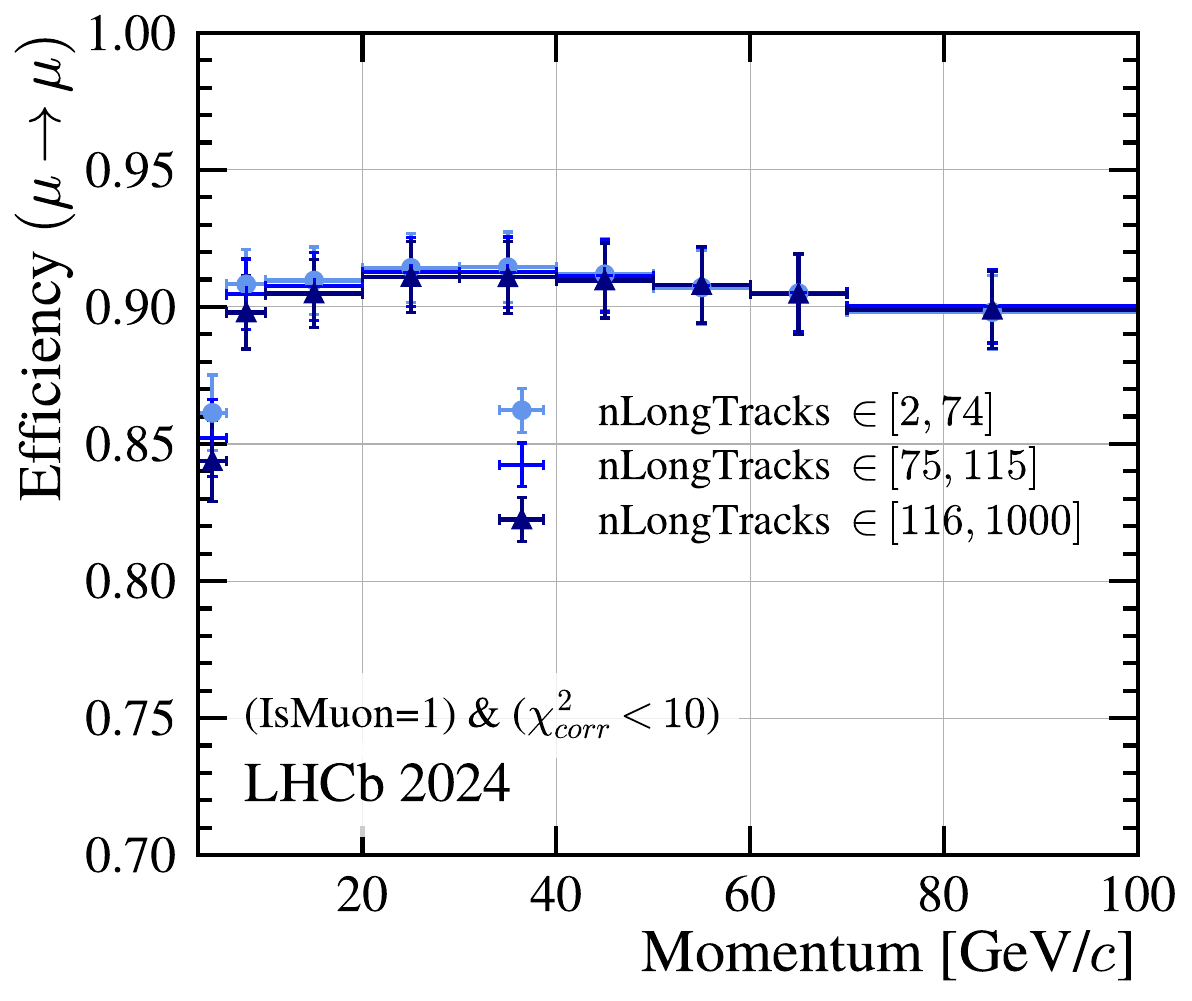}
    \includegraphics[width=0.49\linewidth]{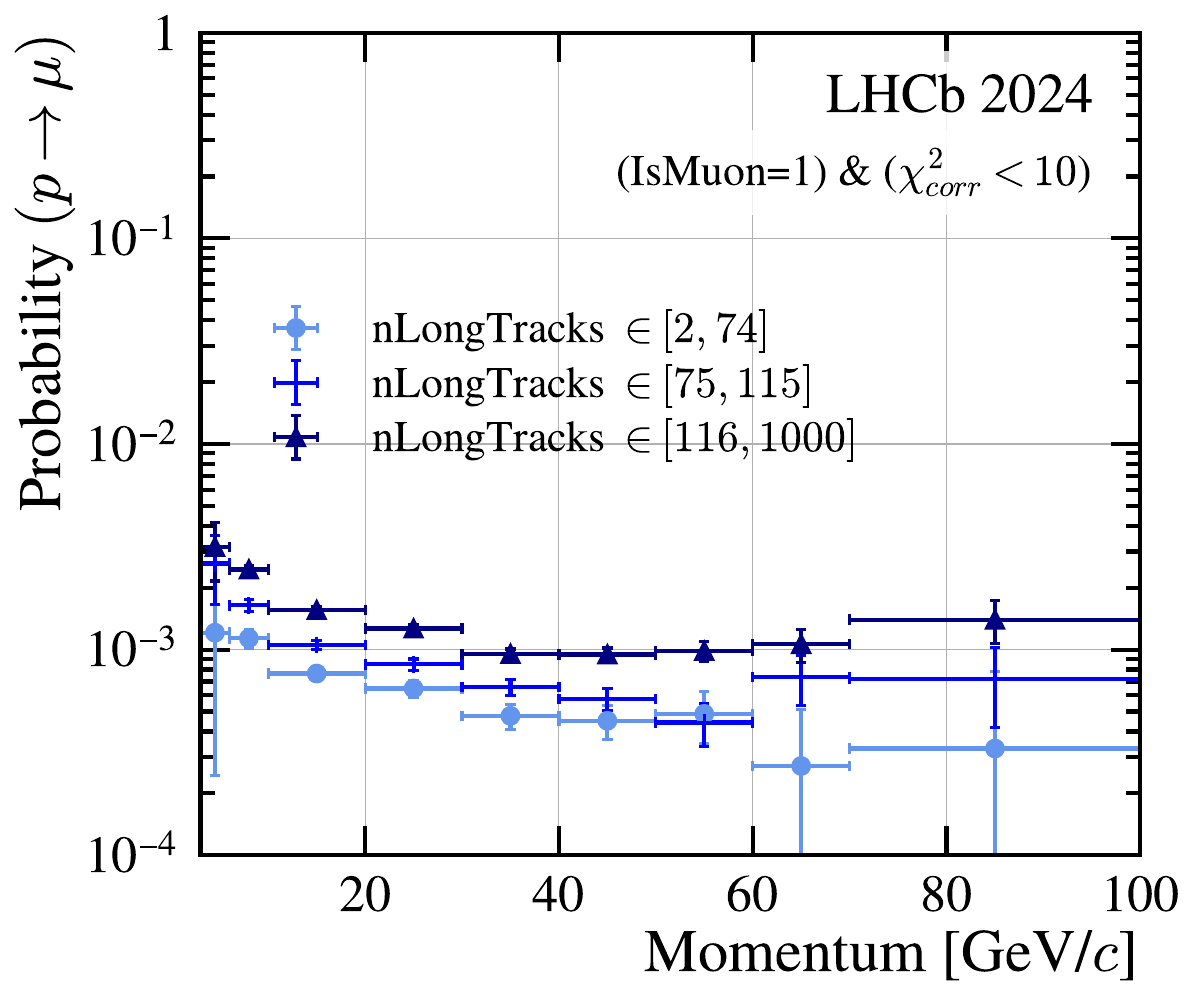}
    
    \caption{Muon efficiency and proton misidentification probability of the $\chi^2_{corr}<10$ selection in three equally populated bins of \texttt{nLongTracks}.}
    \label{fig:misid_mult}
\end{figure}

In Fig.~\ref{fig:misid_mult}, the performance of the muon identification is evaluated as a function of the event occupancy, represented by the number of long tracks in the event (\texttt{nLongTracks}). No significant drop in the muon identification efficiency is observed in crowded events, whereas the proton misidentification probability, driven by combinatorial hits, varies by a factor of two between low- and high-occupancy events. In the $\Lambda \to p\pi^-$ sample, the average number of long tracks per event is 120, whereas the typical Run~2 value was 50.

\begin{figure}[!ht]
    \centering
    
    \includegraphics[width=0.9\linewidth]{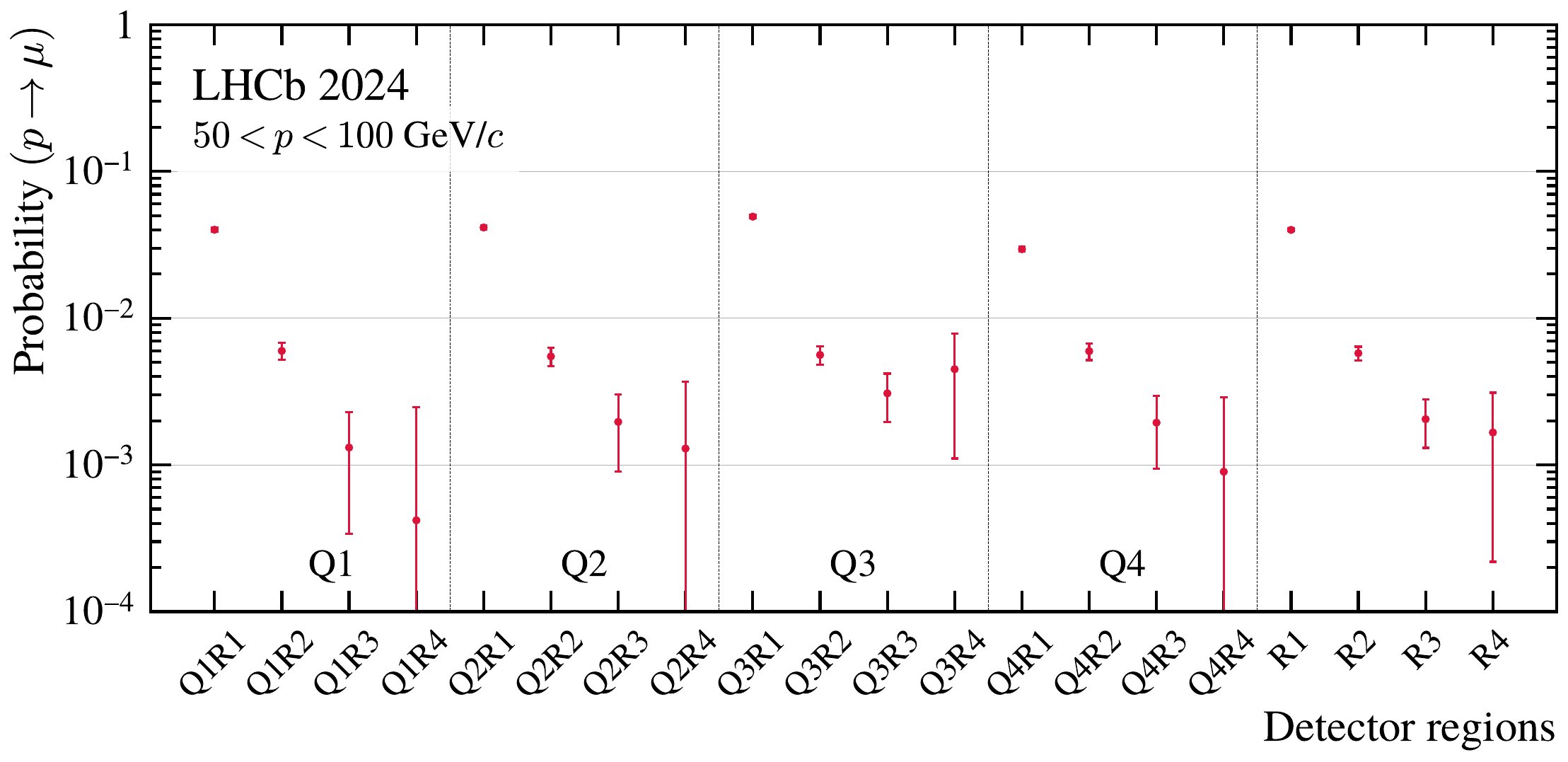}
    
    \caption{Proton misidentification probability for the \texttt{IsMuon} selection across the detector quadrants and regions for tracks with $50<p<100 \gevc$.}
    \label{fig:misid_region}
\end{figure}

The impact of the hit multiplicity can also be observed by computing the misidentification probability of high-momentum protons in each of the four detector regions using the \texttt{IsMuon} selection. Fig.~\ref{fig:misid_region} shows that the misidentification sharply decreases from R1 to R4, following the decrease of the particle flux from the beam pipe outwards. No up--down or left--right asymmetry is observed in the misidentification probability when the sample is further split by detector quadrant.

In conclusion, for tracks having momenta above $10\gevc$, a working point with a muon efficiency around 90\% and a hadron misidentification probability at the per mille level can be reached. This is comparable to the working point observed during Run~2.

\clearpage

\section{Conclusions} 
The MWPCs of the LHCb muon detector operated successfully during Run~1 and Run~2. In Run~3, the same chambers are used, while the readout electronics underwent a complete overhaul. During commissioning, the detector working point was adjusted to operate at five times higher instantaneous luminosity, followed by precise time and spatial calibrations, achieving the required 99\% hit-efficiency target. Within the new fully software-based trigger, a new muon identification algorithm was developed and its performance is evaluated on 2024 calibration data. By exploiting the hit patterns in the muon system, a hadron misidentification probability at the per mille level is achieved with a muon efficiency around 90\%, fulfilling the Upgrade~I detector goals.

\section*{Acknowledgements}

\noindent 

We thank our colleagues in the LHCb collaboration who contributed to the installation and commissioning of the Upgrade I detector, and in particular Federico Alessio and Silvia Gambetta for the review of this document. We thank the technical and administrative staff at the LHCb institutes. We express our gratitude to our colleagues in the CERN accelerator departments for the excellent performance of the LHC. We acknowledge support from CERN and from the national
agencies: INFN (Italy); CNRS/IN2P3 (France); MOST and NSFC (China); STFC (United Kingdom); MICINN (Spain); DOE NP and NSF (U.S.A.). We are indebted to the communities behind the multiple
open-source software packages on which we depend. Individual members have
received support from the European Research Council Starting grant ALPACA 101040710 and the National Science Foundation grant NSF-2310073.


\addcontentsline{toc}{section}{References}
\bibliographystyle{LHCb}
\bibliography{main}

\end{document}

%% file: symbols.tex
\newcommand{\lhcborcid}[1]{\href{https://orcid.org/#1}{\hspace*{0.1em}\raisebox{-0.45ex}{\includegraphics[width=1em]{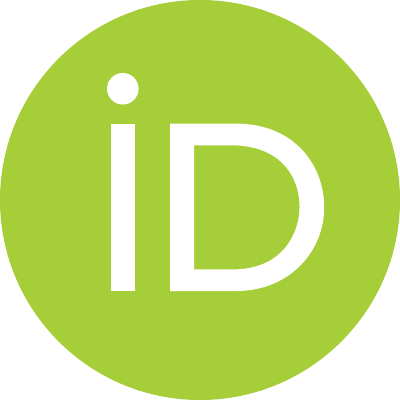}}}}
       
\newcommand{\aunit}[1]{\ensuremath{\text{\,#1}}}   
\newcommand{\gev}{\aunit{Ge\kern -0.1em V}\xspace}
\newcommand{\mev}{\aunit{Me\kern -0.1em V}\xspace}
\newcommand{\mevc}{\ensuremath{\aunit{Me\kern -0.1em V\!/}c}\xspace}
\newcommand{\gevc}{\ensuremath{\aunit{Ge\kern -0.1em V\!/}c}\xspace}
\def\pt         {\ensuremath{p_{\mathrm{T}}}\xspace}
\def\ns   {\ensuremath{\aunit{ns}}\xspace}
\def\mhz  {\ensuremath{\aunit{MHz}}\xspace}
\def\khz  {\ensuremath{\aunit{kHz}}\xspace}
\def\cm   {\aunit{cm}\xspace}

\def\mm   {\aunit{mm}\xspace}
\def\sec  {\ensuremath{\aunit{s}}\xspace}

\def\nb {\aunit{nb}\xspace}
\def\invnb {\ensuremath{\nb^{-1}}\xspace}

\def\fb   {\ensuremath{\aunit{fb}}\xspace}
\def\invfb   {\ensuremath{\fb^{-1}}\xspace}

\def\hr   {\aunit{hr}\xspace}
\newcommand{\ie}{\mbox{\itshape i.e.}\xspace}
\def\Pp      {\ensuremath{\mathrm{p}}\xspace}   
\def\proton      {{\ensuremath{\Pp}}\xspace}
\def\sPlot{\mbox{\em sPlot}\xspace}
\def\lhcb   {\mbox{LHCb}\xspace}

\def\kGy {\aunit{kGy}\xspace}
\def\litre {\aunit{L}\xspace}
\def\volt {\aunit{V}\xspace}
\def\kvolt {\aunit{kV}\xspace}
\def\mC {\aunit{mC}\xspace}

%% file: authors.tex
\begin{center}
P.~Albicocco$^1$\lhcborcid{0000-0001-6430-1038}, 
M.~Anelli$^1$, 
F.~Archilli$^{4,a}$\lhcborcid{0000-0002-1779-6813},
M.~Atzeni$^{16}$\lhcborcid{0000-0002-3208-3336}, 
W.~Baldini$^5$\lhcborcid{0000-0001-7658-8777}, 
A.~Balla$^1$,
S.~Belin$^{13}$\lhcborcid{0000-0001-7154-1304}, 
N.~Bondar$^{15}$\lhcborcid{0000-0003-2714-9879}, 
D.~Brundu$^2$\lhcborcid{0000-0003-4457-5896}, 
S.~Cadeddu$^2$\lhcborcid{0000-0002-7763-500X},
S.~Cal\`i$^8$\lhcborcid{0000-0001-9056-0711},
A.~Cardini$^2$\lhcborcid{0000-0002-6649-0298}, 
M.~Carletti$^1$, 
A.~Casais~Vidal$^{16}$\lhcborcid{0000-0003-0469-2588},
V.~Chulikov$^1$\lhcborcid{0000-0002-7767-9117}, 
A.~Chubykin$^{15}$\lhcborcid{0000-0003-1061-9643}, 
P.~Ciambrone$^1$\lhcborcid{0000-0003-0253-9846}, 
L.~Congedo$^{10}$\lhcborcid{0000-0003-4536-4644}, 
A.~Contu$^2$\lhcborcid{0000-0002-3545-2969}, 
F.~Debernardis$^{10,b}$\lhcborcid{0009-0001-5383-4899}, 
E.~De~Lucia$^1$\lhcborcid{0000-0003-0793-0844}, 
G.~De~Robertis$^{10}$, 
M.~De~Serio$^{10,b}$\lhcborcid{0000-0003-4915-7933}, 
P.~De~Simone$^1$\lhcborcid{0000-0001-9392-2079}, 
F.~Dettori$^{2,c}$\lhcborcid{0000-0003-0256-8663},
L.~Dreyfus$^3$\lhcborcid{0009-0000-2823-5141},
A.~Dzyuba$^{15}$\lhcborcid{0000-0003-3612-3195},
G.~Felici$^{1}$\lhcborcid{0000-0001-8783-6115},
M.~Gatta$^1$\lhcborcid{0000-0002-7179-3023},
A.~Granik$^{15}$,
G.~Graziani$^6$\lhcborcid{0000-0001-8212-846X},
D.~Ilin$^8$\lhcborcid{0000-0001-8771-3115},
S.~Kotriakhova$^{17}$\lhcborcid{0000-0002-1495-0053},
R.~Kristic$^7$,
A.~Lai$^2$\lhcborcid{0000-0003-1633-0496},
R.~Litvinov$^{2,c}$\lhcborcid{0000-0002-4234-435X},
G.~Manca$^{2,c}$\lhcborcid{0000-0003-1960-4413},
F.~Manganella$^{4,a}$\lhcborcid{0009-0003-1124-0974},
S.~Mariani$^7$\lhcborcid{0000-0002-7298-3101},
G.~Martellotti$^9$\lhcborcid{0000-0002-8663-9037},
E.~Minucci$^1$\lhcborcid{0000-0002-3972-6824},
R.~Oldeman$^{2,c}$\lhcborcid{0000-0001-6902-0710}, 
M.~Palutan$^1$\lhcborcid{0000-0001-7052-1360},
L.~Paolucci$ ^{12}$\lhcborcid{0000-0003-0465-2893},
G.~Passaleva$^6$\lhcborcid{0000-0002-8077-8378},
A.~Pastore$^{10}$\lhcborcid{0000-0002-5024-3495},
D.~Pinci$^9$\lhcborcid{0000-0002-7224-9708},
R.~Quagliani$^{7}$\lhcborcid{0000-0002-3632-2453},
T.~Rong$^{11}$\lhcborcid{0000-0002-5479-9212},
R.~Santacesaria$^{9}$\lhcborcid{0000-0003-3826-0329},
M.~Santimaria$^{1,\,\ast}$\lhcborcid{0000-0002-8776-6759},
E.~Santovetti$^{4,a}$\lhcborcid{0000-0002-5605-1662},
A.~Saputi$^{5,7}$\lhcborcid{0000-0001-6067-7863},
C.~Satriano$^{9,d}$\lhcborcid{0000-0002-4976-0460},
A.~Satta$^4$\lhcborcid{0000-0003-2462-913X},
B.~Sciascia$^1$\lhcborcid{0000-0003-0670-006X},
B.~Schmidt$^7$\lhcborcid{0000-0002-8400-1566},
T.~Schneider$^7$,
S.~Simone$^{10,b}$\lhcborcid{0000-0003-3631-8398},
J.~Swallow$^7$\lhcborcid{0000-0002-1521-0911},
R.~Vazquez~Gomez$^{14}$\lhcborcid{0000-0001-5319-1128},
S.~Vecchi$^5$\lhcborcid{0000-0002-4311-3166},
C.~Y.~Yu$^{11}$\lhcborcid{0000-0002-4393-2567},
S.~Zhang$^{11}$\lhcborcid{0009-0007-5868-4512}
\bigskip\\
{\normalfont\itshape\footnotesize
$ ^1$INFN Laboratori Nazionali di Frascati, Frascati, Italy\\
$ ^2$INFN Sezione di Cagliari, Monserrato, Italy\\
$ ^3$Aix Marseille Univ, CNRS/IN2P3, CPPM, Marseille, France\\
$ ^4$INFN Sezione di Roma Tor Vergata, Roma, Italy\\
$ ^5$INFN Sezione di Ferrara, Ferrara, Italy\\
$ ^6$INFN Sezione di Firenze, Firenze, Italy\\
$ ^7$European Organization for Nuclear Research (CERN), Geneva, Switzerland\\
$ ^8$Left the LHCb collaboration more than one year ago\\
$ ^9$INFN Sezione di Roma La Sapienza, Roma, Italy\\
$ ^{10}$INFN Sezione di Bari, Bari, Italy\\
$ ^{11}$ School of Physics State Key Laboratory of Nuclear Physics and Technology, Peking University, Beijing, China\\ 
$ ^{12}$Department of Physics and Astronomy, The University of Manchester, Manchester, United Kingdom\\
$ ^{13}$ Instituto Galego de Física de Altas Enerxías (IGFAE), Universidade de Santiago de Compostela,
Santiago de Compostela, Spain\\
$ ^{14}$  ICCUB, Universitat de Barcelona, Barcelona, Spain\\ 
$ ^{15}$  Affiliated with an institute formerly covered by a cooperation agreement with CERN\\ 
$ ^{16}$ Massachusetts Institute of Technology, Cambridge, MA, United States \\
$ ^{17}$ Institute of Nuclear Physics, Almaty, Republic of Kazakhstan\\
$ ^a$ Università di Roma Tor Vergata, Roma, Italy\\
$ ^b$ Università di Bari, Bari, Italy\\
$ ^c$ Università di Cagliari, Cagliari, Italy\\
$ ^d$ Università della Basilicata, Potenza, Italy\\
}
\bigskip

$^{\ast}$\textit{Corresponding author:} 
\href{mailto:marco.santimaria@cern.ch}{marco.santimaria@cern.ch}

\end{center}
\vspace{\fill}

%% file: tab1.tex
\begin{table}[tb]
    \centering
    \small
    \begin{tabular}{ccccc}
        \toprule
\textbf{\makecell{Station\\and region}} & $x$ [cm] & $y$ [cm] & \makecell{$Q_{\rm{int}}$ (2022--2025) \\ {[mC/cm]}} & \makecell{$Q_{\rm{int}}$ (2022--2033) \\ {[mC/cm]}} \\
\midrule
        M2R1 & 30 & 25 & 226 & 797 \\
        M2R2 & 60 & 25 & 173 & 610 \\
        M2R3 & 120 & 25 & 33 & 115 \\
        M2R4 & 120 & 25 & 4 & 15 \\
\midrule
        M3R1 & 32 & 27 & 92 & 326 \\
        M3R2 & 65 & 27 & 39 & 138 \\
        M3R3 & 130 & 27 & 5 & 17 \\
        M3R4 & 130 & 27 & 1.3 & 4 \\
\midrule
        M4R1 & 35 & 29 & 83 & 292 \\
        M4R2 & 70 & 29 & 20 & 69 \\
        M4R3 & 139 & 29 & 4 & 13 \\
        M4R4 & 139 & 29 & 0.4 & 1.3 \\
\midrule
        M5R1 & 37 & 31 & 96 & 338 \\
        M5R2 & 74 & 31 & 21 & 75 \\
        M5R3 & 149 & 31 & 11 & 38 \\
        M5R4 & 149 & 31 & 3 & 11 \\
        \bottomrule
    \end{tabular}
    \caption{Integrated charge for the most irradiated chambers in each station and region, measured for the period 2022--2025 and expected for the period 2022--2033. The dimensions of the chambers in each region are also provided, while the wire spacing is 2 mm everywhere.}
    \label{tab:1}
\end{table}

%% file: tab2.tex
\begin{table}[tb]
        \centering
        \small
        \begin{tabular}{cc}
            \toprule
            \textbf{Momentum range [GeV/\boldmath{$c$}]} & \textbf{Required hits} \\
            \midrule
            $3 < p < 6$    & M2 \& M3 \\
            $6 < p < 10$   & M2 \& M3 \& (M4 $\lor$ M5) \\
            $p > 10$       & M2 \& M3 \& M4 \& M5 \\
            \bottomrule
        \end{tabular}
        \caption{Hits required in the FOI for each muon station to satisfy the \texttt{IsMuon} condition.}
        \label{tab:2}
    \end{table}